\documentclass[12pt]{article}

\usepackage{lscape}
\usepackage{multirow}
\usepackage{amssymb,amsmath}
\usepackage{amstext}
\usepackage{amscd}
\usepackage{graphics}
\usepackage{epsfig}

\def\wt{\widetilde}
\newcommand{\Rho}{{\mbox{\sf P}}}
\newcommand{\sRho}{\scalebox{.7}{{\mbox{\sf P}}}}

\def\ts{\textstyle}

\def\be{\begin{equation}}
\def\ee{\end{equation}}
\def\beq{\begin{equation}}
\def\eeq{\end{equation}}
\def\bea{\begin{eqnarray}}
\def\eea{\end{eqnarray}} 

\def\eqn#1{(\ref{#1})}

\def\nn{\nonumber}

\def\sideremark#1{\ifvmode\leavevmode\fi\vadjust{\vbox to0pt{\vss
 \hbox to 0pt{\hskip\hsize\hskip1em
 \vbox{\hsize3cm\tiny\raggedright\pretolerance10000
  \noindent #1\hfill}\hss}\vbox to8pt{\vfil}\vss}}}

\begin{document}
\thispagestyle{empty}

\vspace{.8cm}
\setcounter{footnote}{0}
\begin{center}
\vspace{-25mm}
{\Large
 {\bf Tractors, Mass, and Weyl Invariance}\\[5mm]

 {\sc \small
     A.R.~Gover$^{\mathfrak G}$, A.~Shaukat$^{\mathfrak S}$,  and A.~Waldron$^{\mathfrak W}$\\[4mm]

            {\em\small ${}^{\mathfrak G}\!$
            Department of Mathematics\\ 
           The University of Auckland,
            Auckland, New Zealand\\
            {\tt gover@math.auckland.ac.nz}\\[1mm]  
            ${}^{\mathfrak S}\!$
            Physics Department\\  
            University of California,
            Davis CA 95616, USA\\
            {\tt ashaukat@ucdavis.edu}
            \\[1mm] 
            ${}^{\mathfrak W}\!$
            Department of Mathematics\\ 
            University of California,
            Davis CA 95616, USA\\[-2mm]
            {\tt wally@math.ucdavis.edu}
            }}

 }

\bigskip

{\sc Abstract}\\[-4mm]
\end{center}

{\small
\begin{quote}

Deser and Nepomechie established a relationship between masslessness
and rigid conformal invariance by coupling to a background metric and
demanding local Weyl invariance, a method which applies neither to
massive theories nor theories which rely upon gauge invariances for
masslessness.  We extend this method to describe massive and gauge
invariant theories using Weyl invariance.  The key idea is to
introduce a new scalar field which is constant when evaluated at the
scale corresponding to the metric of physical interest. This technique
relies on being able to efficiently construct Weyl invariant theories.
This is achieved using tractor calculus--a mathematical machinery
designed for the study of conformal geometry.  From a physics
standpoint, this amounts to arranging fields in multiplets with
respect to the conformal group but with novel Weyl transformation
laws. Our approach gives a mechanism for generating masses from Weyl
weights. Breitenlohner--Freedman stability bounds for Anti de Sitter
theories arise naturally as do direct derivations of the novel Weyl
invariant theories given by Deser and Nepomechie. In constant
curvature spaces, partially massless theories---which rely on the
interplay between mass and gauge invariance---are also generated by
our method. Another simple consequence is conformal invariance of the
maximal depth partially massless theories.  Detailed examples for
spins~$s\leq 2$ are given including tractor and component actions,
on-shell and off-shell approaches and gauge invariances. For all spins
$s\geq2$ we give tractor equations of motion unifying massive,
massless, and partially massless theories.

\bigskip

\bigskip

\end{quote}
}

\newpage


\section{Introduction}

The history of Weyl invariance~\cite{Weyl:1929fm} as a principle for developing physical theories is a long one. Notable early examples include Dirac's formulation
of conformally invariant four-dimensional wave equations in six dimensions~\cite{Dirac} and Zumino's work relating Weyl transformations to the conformal group~\cite{Zumino} and the introduction of Weyl compensator fields by Deser and Zumino~\cite{Zumino,Deser0}.
We pick up the story with the investigations of Deser and Nepomechie in the early 1980's who found various novel conformally invariant theories
in a study of masslessness in constant curvature spaces~\cite{Deser:1983tm}. These included a conformal, but not gauge invariant
vector theory; the partially massless spin two theory~\cite{Deser:1983tm,Higuchi:1986py,Deser:2001pe} and a trace-free spin two model~\cite{Drew:1980yk,Barut:1982nj}. Their
criteria for constructing these theories was to require rigid conformal invariance and then use the fact
that constant curvature spaces are conformally flat to establish lightlike propagation. To construct conformally invariant
theories\footnote{By now, at least in $d$-dimensional Minkowski spaces, a general analysis of conformally invariant wave equations has been given~\cite{Shaynkman:2004vu}. Interestingly enough, that approach relies on a study of ${\mathfrak o}(d+2)$ modules. Other recent approaches to conformally invariant wave equations 
include~\cite{Marnelius:2008er,Arvidsson:2006fq}.} 
in constant curvature spaces, a trick was employed: first one couples the model 
to an arbitrary (conformally flat) metric and requires that this system be Weyl invariant (where, of course, both the fields and
metric transform accordingly under local Weyl transformations). 
This ensures rigid conformal invariance upon specializing to a constant curvature metric when only the fields are transformed.
This clever maneuver underlies a deep relationship between mass and Weyl invariance that we will explore in this article.

Rigid conformal invariance is {\it not} the mechanism underlying masslessness for many theories (for instance, consider four-dimensional gravitons).
More generally gauge invariance implies masslessness but again there are exceptions even to this rule 
(aside from the trivial example of St\"uckelberg gauge invariances~\cite{Stuck}, three-dimensional topologically massive theories~\cite{Deser:1982vy} are gauge invariant but support massive propagation\footnote{Save for special choices of their mass parameters which restore lightlike propagation in constant curvature spaces~\cite{Carlip:2008eq}.}).
Indeed, subsequently to the work of Deser and Nepomechie, higher spin partially massless constant curvature theories were discovered~\cite{Deser:2001pe}. These theories are gauge invariant and in four dimensions they are massless in the sense that propagation is lightlike. 
Yet, at non-maximal depth\footnote{Partially massless theories are characterized by higher derivative gauge invariances, the number of which is called the depth. In fact at maximal depth these theories enjoy a scalar gauge invariance (just as for Maxwell theory), and were proven to be conformal~\cite{Deser:2004ji}. These theories could have been predicted by Deser--Nepomechie methods, but computing the detailed Weyl invariant couplings to a background metric would be extremely arduous. Our approach provides a simple solution to this problem and therefore an alternate proof to that given in~\cite{Deser:2004ji}.} these theories are not conformal, so could not be found using the methods of Deser and Nepomechie. Nonetheless, in this article we show that Weyl invariance can be used to construct not only conformally invariant theories, but also gauge invariant massless ones and even massive and partially massless theories.

A hint that our viewpoint may be correct can be gleaned from the work of Dolan, Nappi and Witten who constructed an AdS/CFT correspondence 
for partially massless theories~\cite{Dolan:2001ih}. In particular, they related partially massless gauge invariances to conformally invariant, higher derivative boundary operators constructed by conformal geometers in the mathematics literature~\cite{East,Bast,Rice,Slov}. Conformal geometry is the study of manifolds equipped with a conformal equivalence class of metrics with equivalence defined by equality modulo Weyl transformations. This hints that Weyl invariance may indeed be  the correct underlying principle.

The question then, is if Weyl invariance is used as
the guiding principle for constructing massive and massless theories, how does one avoid obtaining only conformally invariant theories when specializing to metrics
with conformal isometries? Put simpler; one does not expect to find Weyl invariant massive theories. The solution is to add an additional
Weyl scale to the theory. {\it I.e.}, instead of coupling the model of interest to a background metric, one also introduces an extra scalar field
and asks that theory to be Weyl invariant. This seems counterintuitive, because upon setting the metric to the background metric of interest,
an extra unwanted scalar remains. However, for any metric chosen from a conformal class of metrics there exists a canonical scalar field that is constant
for that choice of metric. Weyl invariance holds only when this scalar, the metric, and underlying physical fields {\it all} transform. Upon choosing 
a given metric, conformal invariance is broken if this scalar is held constant. One may think of this scalar/Weyl scale as measuring the breaking of Weyl 
invariance by adding masses. This is essentially the main idea of the Weyl compensator method~\cite{Zumino} (a very useful pedagogical introduction is~\cite{Siegel}).
As a consequence of this approach, masses are related to Weyl weights (which become conformal weights when specializing
to metrics with conformal isometries).

This still leaves us with the practical problem of systematically and efficiently constructing Weyl invariant theories.
However, the Weyl scale method can be neatly incorporated with tractor calculus methods. This gives an elegant approach to constructing the 
Weyl invariant theories underlying 
massless, massive and partially massive theories and does so within a single framework. The main idea is that while the metric transforms in the standard 
way under Weyl transformations, the physical fields have transformations corresponding to sections of certain ``tractor bundles''. In simple terms,
this means that fields are arranged as multiplets of the conformal group and have local transformations valued in a parabolic subgroup thereof.
We will refer to such multiplets as tractors, importantly there exists  a set of differential operators that map tractors to tractors and facilitate the
construction of Weyl invariant quantities. These ideas are explained in detail in the mathematics literature in~\cite{G,GP,CG}, while original references are~\cite{Thomas,BEG,East,BaG}.
A brief physical {\it expliqu\'e} may be found in~\cite{GW}. The main notations and ideas required here are given in the next section.

Having explained some key ideas of tractor calculus in Section~\ref{tractors}, the rest of this article is devoted to physical examples
starting with scalars followed by a detailed discussion of vector theories that unifies Maxwell's and Proca's equations. Spin two is handled
thereafter, being the first case where partially massless theories appear. Section~\ref{s>2} is devoted to higher spin theories. Many of
our calculations are performed in a constant curvature setting for both simplicity and physical reasons, but others are germane to any background, as are 
the methods presented here. Although our main aim is to  develop theories in fixed backgrounds, the metric can be made dynamical in a simple way.
This and the theory of Weyl compensators is discussed in Section~\ref{Compensate}.
In our Conclusions we speculate about Weyl invariant ancestors of three-dimensional topologically massive
theories as well as new methods for computing massive spin two interactions. We also include Appendices with detailed formul\ae\ to assist 
readers convert tractor quantities to their more familiar tensor components and review an elegant index-free algebraic technique for handling higher spin
computations~\cite{Damour:1987vm,Hallowell:2005np,Hallowell:2007zb}.

\section{The Tractor Philosophy}

\label{tractors}

Under (local) Weyl transformations the metric transforms as \be
g_{\mu\nu} \mapsto \Omega^{2} g_{\mu\nu}\, .\label{Metric} \ee Let us
denote \be \Upsilon_\mu=\Omega^{-1}\partial_\mu\Omega\, .  \ee In
conformal geometry (in dimensions $d\geq 3$) the trace adjusted
version of the Ricci tensor, plays an important {\it r\^ole}; this is
the $rho$-tensor (sometimes called the Schouten tensor)
$\Rho_{\mu\nu}$, \be \Rho_{\mu\nu}=\frac1{d-2}\,
\Big(R_{\mu\nu}-\frac12\frac1{d-1}\, g_{\mu\nu}\, R\Big)\, ,\qquad
\Rho\equiv \Rho^\mu{}_\mu = \frac{R}{2(d-1)}\, .  \ee Firstly the
Riemann tensor can be simply expressed~as \be
R_{\mu\nu\rho\sigma}-W_{\mu\nu\rho\sigma}=\Rho_{\mu\rho}g_{\nu\sigma}
-\Rho_{\nu\rho}g_{\mu\sigma} -\Rho_{\mu\sigma}g_{\nu\rho}
+\Rho_{\nu\sigma}g_{\mu\rho}\, , \ee where $W_{\mu\nu\rho\sigma}$ is
the trace-free Weyl tensor--the obstruction to conformal flatness in
dimensions~$d>3$.  In three dimensions the Weyl tensor vanishes and
instead, the Cotton tensor, which is the curl of the
$rho$-tensor, measures the failure of conformal flatness.  However,
of far more importance is that the vielbein, Levi-Civita connection
and $rho$-tensor can be combined into a larger ``tractor''
connection\footnote{
  It is interesting to briefly recount the history of this connection
  in the physics literature. There it was first encountered in a study
  of conformal gravity~\cite{Kaku:1978nz} undertaken as a stepping
  stone to theories of conformal supergravity. Indeed this approach is
  part of a general program for gauging spacetime
  algebras~\cite{Ferrara:1977ij,Townsend:1979ki,PvN}.  A related
  description of conformally invariant field theories relying on
  ``conformal space''~\cite{Klein,Weyl1,Kastrup,Dirac} may be found
  in~\cite{Preitschopf:1998ei}.}  
 \be {\cal A}_\mu=
\begin{pmatrix} 0&-e_{\mu n}&0\\[2mm]
\Rho_{\mu}{}^m&\omega_\mu{}^m{}_n&e_{\mu}{}^m\\[3mm]
0&-\Rho_{\mu n}&0
\end{pmatrix}\, .\label{calA}
\ee
Evidently, on Lorentzian signature
manifolds\footnote{Throughout this article we work in Lorentzian
  signature, although essentially every equation we display is valid
  in arbitrary metric signatures.},  this is a Weyl invariant $so(d,2)$-connection. In the component presentation we have here, this is captured
by the Weyl transformation property 
\be {\cal A}_\mu\mapsto U(\partial_\mu U^{-1}) + U
    {\cal A}_\mu U^{-1}\, , 
\ee 
when the metric is transformed according to \eqn{Metric}.
Here the $SO(d,2)$-valued matrix $U$
    is given by\footnote{We denote curved indices $\mu,\nu,\ldots$ and
      flat indices $m,n,\ldots$.}  \be U\equiv U(\Omega)=
\begin{pmatrix}
\Omega&0&\;0\;\\[2mm]
\Upsilon^m&\delta^m_n&\;0\;\\[2mm]
-\frac12\Omega^{-1}\,\Upsilon_r\Upsilon^r&-\Omega^{-1}\Upsilon_n&\Omega^{-1}
\end{pmatrix}\, ,
\label{Um}
\ee

Notice that $U(\Omega)$ above actually takes values in a parabolic subgroup of
$SO(d,2)$.  In tractor calculus one studies multiplets with gauge
transformations given by the matrix $U(\Omega)$, or, in tighter
language, sections of so-called tractor bundles with parallel transport defined
by the connection ${\cal A}_{\mu}$. For example a tractor vector field of
weight\footnote{The term ``conformal weight'', or just
  ``weight'' is often used in mathematics literature where
  Weyl invariance is often called conformal invariance. 
  } 
  $w$ is
a system consisting of functions $T^+$, $T^-$ and a vector field $T^m$ that is required to satisfy the Weyl transformation law 
\bea T^M\equiv
\begin{pmatrix}
T^+\\T^m\\T^-
\end{pmatrix}&\mapsto&\Omega^w U^M{}_N T^N = \Omega^w 
\begin{pmatrix}
\Omega T^+\\
T^m + \Upsilon^m T^+\\
\Omega^{-1}[T^--\Upsilon_n T^n -\frac12 \Upsilon_n\Upsilon^n T^+]
\end{pmatrix}\, , \nn\\\label{tractorvector}
\eea 
in response to~\eqn{Metric}.  We often say the tractor components
$T^+$, $T^m$ and $T^-$ are placed in the the top, middle and bottom
slots, respectively.  The generalization of the above transformation
rule to tractor tensors is the canonical one. On weight zero tractor vector fields the covariant  gradient operator, is given by
$$
{\cal D}_\mu = \nabla_\mu + \widehat{\cal A}_\mu\, ,
$$ where $\nabla_\mu$ is the Levi-Civita gradient operator acting on
the three slots and~$\widehat{\cal A}$ denotes the matrix in~\eqn{calA}
but without  $\omega$ in its middle slot. 
 This formula then extends to give a covariant
gradient operator on arbitrary tractor fields via the Leibniz rule.
(In fact, when the metric $g_{\mu\nu}$ is fixed, we also use this formula for
fields which are tensor products of tractor fields with tensors.)  It
follows immediately\footnote{One uses the 
  identity $U^M{}_R{}U^N{}_S\eta^{RS}=\eta^{MN}$ where
  $U$ is given in~\eqn{Um}.}  that the tractor metric 
\be
\eta_{MN}=\begin{pmatrix}0&0&1\\0&\eta_{mn}&0\\1&0&0\end{pmatrix}\,
,\label{eta} \ee 
is a weight zero, symmetric, rank two tractor tensor
that is parallel with respect to ${\cal D}_\mu$.  Using the last fact
it is safe in calculations to use the tractor metric and its inverse
to raise and lower tractor indices $M,N,\ldots$ in the usual fashion,
and this we shall do without further mention. Along similar lines, it
is easy to see that 
\be
X^M\equiv\begin{pmatrix}\ 0\ \\0\\1\end{pmatrix} 
\ee 
is a (Weyl
invariant) weight one tractor vector. Note that, by contraction, $X^M$
may be used to ``project out'' the top slot of a tractor vector field.

The Thomas $D$-operator acts on weight $w$ tractors by 
\be
D^M\equiv\begin{pmatrix}(d+2w-2)w\\[2mm](d+2w-2){\cal
  D}^m\\[2mm]-({\cal D}_\nu{\cal D}^\nu + w P)\end{pmatrix}\label{D}
\ee 
and is extremely important since it allows us to produce new,
weight $w-1$, tractors from old.  It is important not to confuse the
symbol $D^M$ with the covariant gradient operator ${\cal D}^m$
(contracted with an inverse vielbein) which appears in the middle slot
of the Thomas $D$-operator. Although $D^M$ is {\it not} a covariant
derivative, nevertheless it can often be employed to similar
effect. Let us denote the Laplacian on scalars by $\Delta \equiv {\cal
  D}_\nu {\cal D}^\nu$.  Acting on a weight $w=1-\frac d2$ scalar
$\varphi$, the Thomas $D$-operator yields, in particular, 
\be D^M
\varphi= \begin{pmatrix}0\\[1mm]0\\[1mm]-\Big(\Delta - \frac{d-2}2 \,
  \Rho\Big)\ \varphi\end{pmatrix}\, , 
\ee 
where the operator in the bottom slot is the conformally invariant
wave operator (or in Riemannian signature, the Yamabe operator).
Hence~$D^M\!\!$~$\varphi$~$=$~$0$ yields the equation of motion of a conformally
improved scalar $\Delta \varphi = \frac{d-2}2\, \Rho \varphi$.  It is
also important to note that $D^M$ reduces the weight of a tractor by
one and that \be D^M D_M = 0\, .\label{null} \ee

In this article, we not only construct tractor equations of motion,
but also action principles.  To that end we need to know how to
integrate the Thomas $D$-operator by parts. Notice that the
$D$-operator does not satisfy a Leibniz rule; this is to be expected
because for example its bottom slot is a second order differential
operator.  Nonetheless, at commensurate weights, an integration by
parts formula does hold. For example, if $V^M$ is a weight $w$ tractor
vector and $\varphi$ is a weight $1-d-w$ scalar (with compact support) then (see {\it e.g.} \cite{BrGononlocal})
\be \int \sqrt{-g} \, V_M D^M \varphi = \int \sqrt{-g}\,
\varphi D^M V_M\, .\label{parts} 
\ee 
Notice, there is no sign flip in
this formula and that the integral itself is Weyl invariant because
the metric determinant carries Weyl weight~$d$. An analogous formula
holds for tractor tensors.

The final piece of tractor technology we will need is how to
handle choices of scale without losing contact with the tractor systems. 
Consider a
conformal class of metrics $[g_{\mu\nu}]$ with equivalence defined by Weyl transformations~\eqn{Metric}; so \be
[g_{\mu\nu}]=[\Omega^2 g_{\mu\nu}]\, .
\ee
Adjoining  the conformal class of a weight $w=1$, non-vanishing, scalar $\sigma$ to this
with equivalence 
\be
[g_{\mu\nu},\sigma]=[\Omega^2 g_{\mu\nu},\Omega \sigma]\, ,
\ee
we can use $\sigma$ to uniquely (up to an overall constant factor)
pick a metric $g_{\mu\nu}^0$ from this equivalence class by requiring
the accompanying representative scalar $\sigma^0$ is constant for that
choice. 
Evaluated at that choice of metric, the Thomas $D$-operator
acting on $\sigma^0$ is simple and plays a distinguished {\it
  r\^ole}. Therefore (dropping the superscript ``0'') we
define the weight zero tractor vector\footnote{It is important to
  realize this is the expression for $I^M$ evaluated at a special
  scale.
To obtain its correct tractor transformation law, one must
  first transform $\sigma$ as a weight one scalar and subsequently
  evaluate the derivatives in $D^M$.}
\be
I^M=\frac 1d\ D^M \sigma = \begin{pmatrix}
\sigma \\[1mm]0 \\[1mm] -\frac 1 d  \Rho  \sigma\, \end{pmatrix} \, ,\label{I}
\ee
that we term the {\em scale tractor}. 
Notice that
\be
{\cal D}_{\mu} I^M = \sigma \begin{pmatrix}0\\[2mm]\Rho_\mu{}^m-\frac 1 d\, e_\mu{}^m\Rho \\[2mm] -\frac 1 d \,  \partial_\mu \Rho \end{pmatrix}
\ee
so that $I^M$ is parallel if and only if $g_{\mu\nu}$ is an Einstein
metric\footnote{The space of solutions to the requirement of parallel
  $I^M$ can be enhanced to include almost Einstein structures by
  allowing zeroes in $\sigma$.  at conformal
  infinities~\cite{GoPrague,almost}.}. Hence, at arbitrary scales, when
$g_{\mu\nu}$ is conformally Einstein, $D^M\sigma$ is a parallel
tractor.  It also follows that $[D^M,I^N]=0$. If in addition to $I^M$
parallel, one has vanishing Weyl tensor, then $g_{\mu\nu}$ is
conformally flat and it follows that Thomas $D$-operators
commute. Many of the computations in this article pertain to arbitrary
conformal classes of metrics, but as the spin of the systems we study
increases, we typically restrict ourselves to conformally Einstein, or
even conformally flat metrics.

We have now assembled enough tractor technology to construct physical
models. We refer the reader to the
literature~\cite{BEG,G,GP,CG,CGtr,GW} for a more detailed and
motivated account of tractors.

\section{Scalars}

Let us show how to describe a massive scalar field in a curved
background using tractors. In particular we want to exhibit how the
choice of scale can be used to write down a Weyl invariant theory.  Of
course once a choice of scale is included, then there are trivial ways
to form Weyl invariants and Weyl invariant equations. In contrast we
shall be very restrictive in the way the scale and its derivatives may
couple  with matter fields; the scale is only allowed to enter via
the scale tractor $I$. We might view $I$ as breaking conformal
symmetry, but alternatively we can equally view this as a conformal
theory that couples scale via $I$ and this is the notion of {\em Weyl
  invariance} that we shall describe here.  This yields a theory whose
Weyl weights are related to masses.

 Let $\varphi$ be a
weight $w$ scalar field so that under Weyl transformations 
\be \varphi
\mapsto \Omega^w \varphi\, .  \ee 
We propose the Weyl covariant
equation of motion \be I_M D^M \varphi = 0\, .\label{sceom} 
\ee
Importantly, covariance is achieved only upon transforming the metric
$g_{\mu\nu}$, the scale $\sigma$ and the physical field $\varphi$ 
simultaneously. 
 It is convenient to choose the scale such that it is 
constant for the background
metric of interest. In that case $I^M$ takes the simple form~\eqn{I},
so it is easy to use the expressions~\eqn{D} and~\eqn{eta} to evaluate
the ``component'' expression\footnote{We also refer the reader to
  Appendix~\ref{tractor components} for a tractor/tensor component
  dictionary.} for the equation of motion~\eqn{sceom}. We obtain 
(cf.\ \cite{GDN})
\be
-\sigma\Big(\Delta+\frac{2\Rho}{d} w(w+d-1)\Big)\varphi=0\,
.\label{scomp} \ee Our first observation is that when $w=1-\frac d2$,
this is the equation of motion for a conformally improved scalar field
\be \Big(\Delta-\frac R4\, \frac{d-2}{d-1}\, \Big)\varphi=0\, .  \ee
This equation is Weyl covariant so we expect no choice of scale
$\sigma$ to be needed. Indeed this is true, since at this weight, the
equation of motion~\eqn{sceom} can be rewritten as \be D^M \varphi =
0\, ,\label{spsc} \ee because the top and middle slots of the Thomas
$D$-operator vanish. The same conclusion follows from a Weyl invariant
action principle\footnote{Recall that the metric determinant has
  Weyl weight $d$.} for the equation of motion~\eqn{sceom} \be
S[g_{\mu\nu},\sigma,\varphi]=\frac12 \int \sqrt{-g} \, \sigma^{1-d-2w}
\varphi I_M D^M \varphi =
S[\Omega^2g_{\mu\nu},\Omega\sigma,\Omega^w\varphi]\, , \ee because $S$
is independent of $\sigma$ when $w=1-\frac d2$.

For generic $w\neq 1-\frac d2$ we can treat the equation of
motion~\eqn{sceom} as one for a massive scalar by (locally) taking the metric
$g_{\mu\nu}$ to be a representative in the conformal class with constant scalar curvature, assuming there is one.\footnote{In Riemannian signature there is always such a metric. 
If, on the other hand, backgrounds
  with non-constant scalar curvature are desired, we can always make a
  Weyl transformation that brings the equation~\eqn{scomp} to the form
$$
\Big(\Delta+\frac{2\Rho}{d} w(w+d-1)\Big)\varphi = (d+2w-2)\Big( \upsilon^\mu \nabla_\mu -w(\nabla_\mu \upsilon^\mu + \upsilon_\mu\upsilon^\mu)\Big)\varphi\, ,
$$
where $\Rho$ is constant and $\upsilon_\mu = \sigma^{-1}\partial_\mu\sigma$.
} 
Then, the
equation of motion~\eqn{scomp} describes a massive scalar field
\be
(\Delta-m^2)\varphi=0\, ,
\ee
with the mass-Weyl weight relation
\be
m^2 = \frac{2\Rho}{d}\Big[\Big(\frac{d-1}2\Big)^2-\Big(w+\frac{d-1}2\Big)^2\Big]\, .\label{massweight}
\ee
Hence, in summary, the Weyl covariant equation of motion $I_M D^M\varphi =0$ describes massive propagation for generic $w$
and at the conformal value $w=1-\frac d2$ the scale $\sigma$ decouples and we may impose $D^M\varphi=0$.

\begin{figure}
\begin{center}
\epsfig{file=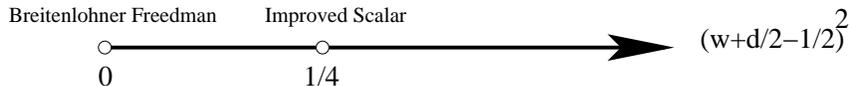,height=1.1cm}
\end{center}
\caption{ The Weyl weight $w$ can be reinterpreted as a scalar mass parameter. Generic values of $w$ (the thick line) give
massive theories, while $w=\frac12 -\frac d2$ and $w=1-\frac d2$ describe a scalar saturating the Breitenlohner--Freedman bound
(in Anti de Sitter space) and an improved scalar, respectively.\label{scalarmass}}
\end{figure}

Finally to set notations, note that for the weight $w=1-\frac{d}{2}$
we have 
\be 
m^2=\frac{\Rho}{2}\, (d-2) = \frac{R}4 \frac{d-2}{d-1}\, .
\ee
Also, in constant curvature backgrounds
\be
R_{\mu\nu\rho\sigma}=\frac{4\Rho}{d}\, g_{\mu[\rho}g_{\sigma]\nu}\, ,
\ee 
and the cosmological constant is related to the $rho$-tensor by
\be \Rho=\frac{R}{2(d-1)}=\frac{\Lambda d}{(d-1)(d-2)}\, , 
\ee 
which
is negative in Anti de Sitter and positive in de Sitter spaces.  The
Breitenlohner--Freedman
bound~\cite{Breitenlohner:1982jf,Mezincescu:1984ev} for stable scalar
propagation in Anti de Sitter space is 
\be m^2\geq \frac{\Rho}{2d}\,
(d-1)^2\, , 
\ee 
and is saturated by setting the second term
in~\eqn{massweight} to zero, so that $w=\frac12 -\frac d2$. It is also
worth observing that, according to \eqn{massweight}, any real weight $w$ obeys this bound.
Figure~\ref{scalarmass} depicts the physical interpretation of the
various values of the Weyl weight~$w$.

\section{Vectors}

\label{vectors}

To describe theories of a vector field $V_\mu$ according to the
tractor philosophy espoused in section~\ref{tractors}, we need to
arrange fields as $SO(d,2)$ multiplets transforming as tractors under
Weyl transformations. This necessitates the addition of auxiliary
fields in order to build a tractor from $V_\mu$. So to begin with we
introduce a weight $w$ tractor vector 
\be V^M=\begin{pmatrix}V^+\\ V^m
\\ V^-\end{pmatrix}\, , 
\ee 
and identify\footnote{Notice that the
  vielbein transforms as $e_\mu{}^m\mapsto \Omega e_\mu{}^m$.}  \be
V_\mu = e_\mu{}^m V_m\, .  \ee This forces us to incorporate, in our
model, the two auxiliary fields $V^\pm$.  With $V_\mu$ these have Weyl
transformation laws (see~\eqn{tractorvector}) 
\bea
V^+&\mapsto&\Omega^{w+1}V^+\, ,\nn\\[2mm] V_\mu\;
&\mapsto&\Omega^{w+1}(V_\mu+\Upsilon_\mu V^+)\, ,\nn\\[1mm]
V^-&\mapsto&\Omega^{w-1}(V^- -\Upsilon^\mu [V_\mu +\frac 12
  \Upsilon_\mu V^+])\, .\label{vecfo} 
\eea 
These formul\ae\ are
perhaps somewhat mysterious from a physical perspective. Firstly, what
do the extra fields $V^\pm$ mean? Secondly, although nothing prevents
us from choosing strange looking Weyl transformations (so long as we
maintain the standard formula for the metric~\eqn{Metric}), these are
certainly non-standard. For example, in four dimensions, Maxwell
theory is Weyl invariant and only requires a single vector field which
is inert under Weyl transformations.  In other dimensions, Deser and
Nepomechie have written down a non-gauge invariant, yet Weyl invariant
vector theory~\cite{Deser:1983tm} which again involves only a single
vector field $V_\mu$, but with weight $w=1-\frac d2$.

Let us hint at a solution to this problem: For special weights the auxiliaries will decouple and can be
set to zero consistently; for generic weights, one auxiliary remains and  will be a St\"uckelberg field enabling us to deal with massive fields in a gauge invariant way.

Since the tractor vector $V^M$ seems to have a too large field content, we may search for Weyl-covariant constraints, using the tractor operators
of section~\ref{tractors}. Hence we impose the requirement
\be
D^M V_M = 0\, .\label{DV}
\ee
(One might also consider the further constraint $X^M V_M=0$ which forces $V^+=0$;  we choose not to at this juncture,
but will encounter it later for special choices of $w$.) This constraint has the advantage that it can be solved algebraically
for $V^-$. In components the solution is\footnote{We often denote contraction of $d$-dimensional indices by a dot ``.'' while  contractions
of tractor indices are given by a slightly higher dot ``$\cdot$''.}
\be
V^M = \begin{pmatrix}V^+\\[2mm]
V^m\\[2mm]
-\frac{1}{d+w-1} \left(\nabla.V -\frac 1{d+2w}\, \Big[\Delta -(d+w-1)\Rho\Big] V^+\right)
\end{pmatrix}\, ,
\ee
at least for $w\neq -\frac{d}{2}, 1-d$. At the weight $-d/2 $ the
equation \eqn{DV} amounts to a conformal wave equation on $V^+$; we
shall discuss this below. 
(As a check of this computation, one can readily verify that the bottom slot of this tractor transforms according to~\eqn{vecfo}
through its dependence on the middle and top slots.) 

Having removed the auxiliary $V^-$ as an independent field, we still need to deal with the auxiliary $V^+$; here we employ 
the standard physics principle~---~gauge invariance. Therefore we posit the local invariance
\be
\delta V^M = D^M \xi\, .\label{eee}
\ee
where the parameter $\xi$ is a weight $w+1$ scalar. This invariance 
respects the constraint~\eqn{DV}
because the operator~$D^M$ is ``null'' (see~\eqn{null}).

It is instructive to write these transformations out in components
\bea \delta V^+ &=& (d+2w)(w+1)\, \xi\, ,\nn\\[2mm] \delta V_\mu \,
&=& (d+2w)\nabla_\mu \xi\, .\label{maxgauge} 
\eea 
Notice that we
recover the standard Maxwell type gauge transformation for the
vector~$V_\mu$, while the auxilary $V^+$ is indeed a St\"uckelberg
field since its gauge transformation is a shift symmetry. We must pay
careful attention to the $w$-dependent coefficients of these
transformations in the computations that follow.

Next, we search for gauge invariant (and Weyl covariant) quantities
built from $V^M$. The first is the rather elegant\footnote{We
  editorialize by reiterating that although $D^M$ is {\it not} the
  covariant derivative, many standard formul\ae\ may be mimicked by
  treating it so.} \be {\cal F}^{MN}=D^M V^N - D^N V^M\, , \ee which for obvious reasons
we shall term the {\em tractor Maxwell curvature}, even though it is
not the curvature of a connection.  Its gauge invariance is manifest
because Thomas $D$-operators commute acting on scalars (for any
conformal class of metrics).

Interestingly enough, for $w\neq -1,-\frac{d}{2}$ there is a second gauge invariant quantity built
by subtracting from $V^M$ an appropriate {\em differential splitting
  operator} applied to $V^M$. The meaning of this is made clear by the
explicit formula: \be \wt V^M = V^M - \frac{1}{(d+2w)(w+1)} D^M X\cdot
V = \begin{pmatrix}0 \\[2mm] \wt V^m \\[2mm] -\frac{1}{d+w-1}
  \nabla. \wt V \end{pmatrix}\, ,
\label{gaugeV}
\ee
where the quantity 
\be
\wt V_\mu = V_\mu -\frac 1 {w+1} \nabla_\mu V^+\, ,
\ee
is easily seen to be invariant under~\eqn{maxgauge}. (Note that the replacement $V^M\rightarrow \wt V^M$
leaves the tractor Maxwell curvature unchanged.)

At this point we are done with kinematics and proceed to develop
dynamics for the tractor $V^M$.  Therefore we now introduce a choice
of scale $\sigma$ and pick a metric where $\sigma$ is constant.  We
propose the tractor equation of motion 
\be I_M {\cal F}^{MN}\equiv G^N
= 0.\label{maxeom} 
\ee 
Observe that $I_N G^N\equiv 0$.  Note also that
other possibilities for the equations of motion, such as $X_M {\cal
  F}^{MN}$ and $D_M {\cal F}^{MN}$ are not interesting. The former
gives trivial dynamics, while the latter either vanishes identically
in a conformally flat setting (and yields trivial dynamics in any
setting).

We note that the equation of motion~\eqn{maxeom} mimics and generalizes the scalar equation of motion~\eqn{sceom} since expanding out
the tractor Maxwell curvature yields
\be
I\cdot D\,  V^N - I_M D^N V^M = 0\, .
\ee
Let us now analyze in detail these equations to see what physics they describe.

We begin with special values of the weight $w$. Examining the gauge
transformations~\eqn{maxgauge} we see that $w=-\frac d2$ and $w=-1$
play special {\it r\^oles}. The case $w=-d/2$ is deceptive,
 since it
appears to be a distinguished value.  In fact it actually amounts to
the massive Proca system\footnote{The technical details are as
  follows: Firstly, both $V^+$ and $V_\mu$ are inert under gauge
  transformations~\eqn{eee}.  Moreover $V^-$ decouples both from the
  field constraint and equation of motion $D\cdot V=0=G^M$; in fact if we assume invertibility of
  $(\Delta-\frac\sRho2 )$, then
 it
  can be gauged away using the gauge invariance $\delta V^-=D^-\xi =
  -(\Delta-\frac{\sRho}2)\xi$.  Then $\Rho V^+$ turns out to be proportional to
  $\nabla.V$ }
that
we will find for generic weights $w$.

The case $w=-1$ is far more interesting. The auxiliary field $V^+$ is
inert under the transformations~\eqn{maxgauge}. Or in other words the
tractor quantity $X\cdot V$ is gauge invariant. Hence we may
consistently impose an additional constraint \be X_M V^M =0\, .  \ee
In that case, a component computation of the equation of
motion~\eqn{maxeom} is rather simple, the only non-vanishing component
of the left-hand side is \be G^n = -\sigma \nabla_m F^{mn}\, , \ee
where $F_{\mu\nu}=2\nabla_{[\mu}V_{\nu]}$ is the usual Maxwell
curvature.  So $G^n=0$ is {\it precisely the system of Maxwell's
  equations in vacua!}  This is hardly surprising since at this value
of $w$, along with the additional $X\cdot V$ constraint, we have a
theory of a single vector $V_{\mu}$ along with its usual Maxwell gauge
invariance.  Notice, the tractor technology does not predict Weyl
(co/in)variance of Maxwell's equations in arbitrary dimensions since
the construction does involve choosing a scale. Rather the combination $-\sigma
\nabla_m F^{mn}$ belongs in the middle slot of a weight $-2$ tractor
vector.  Note also that the value $w=-1$ along with the constraints
$X\cdot V=D\cdot V=0$ implies the usual Weyl transformation rule for
the Maxwell potential~$V_\mu$, namely that $V_\mu$ is inert.

It turns out there is a further distinguished value of the weight
$w$. Namely the canonical engineering dimension of a $d$-dimensional
field, namely~$w=1-\frac d2$ (just as for an improved scalar).
 In four
dimensions this value coincides with the $w=-1$ Maxwell one!  To see
that this value is special we compute the components of the tractor
Maxwell curvature subject to the constraint~\eqn{DV}, but at arbitrary
weight and find
\be
{\cal F}^{MN}=
\left(
\scalebox{.75}{\mbox{\begin{tabular}{ccc} $0$ & $(d+2w-2)(w+1) \wt V^n$&$-\frac{(d+2w-2)(w+1)}{d+w-1}\, \nabla.\wt V$ \\[3mm]
${\rm a/s}$& $(d+2w-2) F^{mn}$&
$\nabla_r F^{rm}
-(w\!+\!1)[2\Rho^m_r\wt V^r\! -\!\Rho \wt V^m \!+\!\frac 1{d+w-1} \nabla^m \nabla.\wt V]$
\\[3mm]
{ a/s}&{ a/s}&$0$\end{tabular}}}
\right)\, ,\label{FF}
\ee
Notice that at $w=1-\frac d2$ every component of ${\cal F}^{MN}$ vanishes save for ${\cal F}^{m-}$.
Therefore there is no longer any need to introduce the scale $\sigma$ to obtain the field equations. Just as we did in~\eqn{spsc} for the improved
scalar field, we may simply replace~\eqn{maxeom} by vanishing of the tractor Maxwell curvature
\be
{\cal F}^{MN}=0\, .
\ee
This gives Weyl covariant equations of motion that depend only on the combination $\wt V_\mu$.
Without loss of generality, therefore, we can gauge away the St\"uckelberg field and are left with
a theory of a single vector $A_\mu\equiv \wt V_\mu\lvert_{V^+=0}$, with Weyl transformation law
\be
A_\mu\mapsto \Omega^{-\frac{d-4}2} A_\mu\, .
\ee
Writing out these apparently novel equations explicitly gives
\be
\Delta A_\mu - \frac 4 d \, \nabla^\nu \nabla_\mu A_\nu +\frac{d-4}{d}\Big(2\, \Rho_\mu^\nu A_\nu -\frac {d+2}2\,  \Rho A_\mu\Big)=0\, .
\ee
These equations have in fact been encountered before---they are
precisely the Weyl invariant, but non-gauge invariant vector theory of
Deser and Nepomechie~\cite{Deser:1983tm}. When $d=4$, they revert to
Maxwell's equations. Observe that the intersection of the conditions
$w=-1$ and $w=1-d/2$ is at $d=4$, 
precisely the value when the Maxwell
theory is Weyl invariant. It is rather pleasing that the simple
tractor equation~\eqn{maxeom} directly generates the curvature
couplings required for Weyl invariance.

Having dealt with special weights $w$, we now analyze generic weights
which will correspond to massive vector fields.  For this we
specialize to conformally Einstein metrics and choose a scale $\sigma$
which is constant when $g_{\mu\nu}$ is Einstein so that
\be
\Rho_{\mu\nu}=\frac\Rho d \, g_{\mu\nu} \, ,\qquad \nabla_\mu \Rho=0\, .
\ee
Moreover, since $I^M$ is now parallel it commutes with the Thomas $D$-operator.
This means that (just as for regular Maxwell theory), it is easy to prove a Bianchi identity for the field equation~$G^M$
in~\eqn{maxeom}: Since we deal with fields~$V^M$ subject to~$D\cdot V=0$ and a gauge invariance~$\delta V^M = D^M \xi$,
we expect (and indeed must require) both a constraint and Bianchi identity for the field equation~$G^M$. Indeed this is the case; it is easy to verify that
\be
I\cdot G = 0 = D\cdot G\, .
\ee
It may be that there exist corrections to the equation of motion $G^M$ 
that allow these conditions to be generalized beyond conformally Einstein metrics, but we have not studied this issue in detail.

Let us now examine the component form of our proposed equations of motion~\eqn{maxeom} at arbitrary~$w$.
A fairly simple computation shows that a combination of the components of $G^M$ takes a familar form
\be
G^m - \frac{1}{d+2w-2} \nabla^m G^+ = -\sigma \Big(\nabla_n F^{nm} +\frac{2\Rho}{d}(w+1)(d+w-2)\wt V^m\Big)\equiv {\cal G}^m\, . \label{Proca}
\ee
{\it Massive vector cognescenti will recognize this as the St\"uckelberg formulation of the Proca equation
in a cosmological background!} It is invariant under the gauge transformations~\eqn{maxgauge}
because it only depends on the gauge invariant combination~$\wt V_\mu$.
At $w=-1$, the second, mass, term vanishes (so long as one sets the St\"uckelberg field $V^+=0$ as a gauge invariant constraint)  which could be expected from the gauge invariances~\eqn{maxgauge}. 
The other components of $G^M$ are simply consequences of~\eqn{Proca}.

At generic $w$ we can 
gauge away the auxiliary St\"uckelberg field using~\eqn{maxgauge}, and the divergence of the equation of motion~\eqn{Proca} $\nabla_\mu {\cal G}^\mu=0$ implies $\nabla_\mu V^\mu=0$, so we obtain a wave equation for
a massive, divergence free vector field
\be
\left\{
\begin{array}{l}
\Big(\Delta +{\textstyle\frac{ {\ts 2}\Rho}{{\ts  d}}}\, [w(w+d-1) -1]\Big) V_\mu=0\, ,\\[5mm]
\quad \nabla^\mu V_\mu = 0\, .
\end{array}
\right.
\ee

The mass-squared of the Proca system is usually defined by the coefficient of $\wt V^m$ in~\eqn{Proca}.
Hence, performing some elementary algebra, we read off the mass-Weyl weight relation
\be
m^2=\frac{2\Rho}{d}\Big[\Big(\frac{d-3}{2}\Big)^2-\Big(w+\frac{d-1}{2}\Big)^2\Big]\label{maxmass}
\ee
Firstly note, that this result predicts a Breitenlohner--Freedman bound
\be
m^2\geq\frac{2\Rho}{d}\, \Big(\frac{d-3}{2}\Big)^2\, ,\ee
for the Proca system. 

We also observe, that if instead of the standard definition of the mass given above, we define a parameter $\mu^2$
by the eigenvalue of the (Bochner) Laplacian so that
\be
\Delta V_\mu = \mu^2 V_\mu\, ,
\ee
then the mass-Weyl weight relation for the spin $s=0$ scalar and $s=1$ Proca systems can be unified as
\be
\mu^2=\frac{2\Rho}d \Big[ \Big(\frac{d-1}{2}\Big)^2-\Big(w+\frac{d-1}{2}\Big)^2+s\Big]\, .
\ee
In fact, this result holds at arbitrary spin $s$ (see section~\ref{s>2}).
This result predicts a arbitrary spin Breitenlohner--Freedman bound
\be
\mu^2\geq\frac{2\Rho}{d}\, \Big[\Big(\frac{d-1}{2}\Big)^2+s\Big]\, .\label{BFs}
\ee
We also note that the three-dimensional topologically massive Maxwell system saturates this Breitenlohner--Freedman bound~\cite{Carlip:2008eq}.

The Proca equation~\eqn{Proca} follows from an action principle
$S[g_{\mu\nu},\sigma,V_\mu,V^+]$ which evaluated at a constant
curvature metric and constant scale reads 
\bea 
\label{Sform}
S&=&\int
\frac{\sqrt{-g}}{\sigma^{d+2w-2}}\Big\{ -\frac 1 4
F_{\mu\nu}F^{\mu\nu} +\frac{\sc P}{d}\, (w+1)(d+w-2)\ \wt V^{\mu}\, \wt V_{\mu}
\Big\}\nn\\[3mm] &=&\frac12 \int \frac{\sqrt{-g}}{\sigma^{d+2w-2}}
\Big\{V^+ {\cal G}^- + V^m {\cal G}_m \Big\} \, .\label{s0} \eea 
Here
${\cal G}^m$ is given in~\eqn{Proca}, and yieldss the Proca equation, while
\be {\cal G}^-=-\frac{2\Rho}{d}\frac{(d+w-2)(d+w-1)}{(w+1)(d+2w-2)}\,
G^+\, . \label{GG-} \ee 
The pair of equations of motion ${\cal
  G}^m=0={\cal G}^-$ are those that come from varying the action
\eqn{s0}.  Although it is obviously gauge invariant with respect
to~\eqn{maxgauge}, Weyl invariance of this action is not manifest.
Therefore we construct an equivalent tractor action \be
S[g_{\mu\nu},\sigma,V^M] = \int\sqrt{-g} \, V_M {\cal H}^M\,
,\label{s2} \ee where the weight $-w-d$ tractor ${\cal H}^M$ is given
(at the choice of scale where $\sigma$ is constant) by 
\be {\cal H}^M
= \sigma^{1-d-2w}\begin{pmatrix}0\\[2mm] {\cal G}^m\\[2mm]{\cal
    G}^- \end{pmatrix}\, .\label{H} \ee 
Because
${\cal H}^M$ is a tractor vector of weight $-w-d$ the action enjoys the Weyl
invariance \be S[g_{\mu\nu},\sigma,V^M] =
S[\Omega^2g_{\mu\nu},\Omega\sigma,\Omega^wU^M{}_NV^N] \, .\label{s1}
\ee Moreover, the integration by parts formula~\eqn{parts}, along with
the invariance of \eqn{s0} (and hence \eqn{H}) with respect to the
gauge transformation \eqn{maxgauge}, implies the Bianchi identity \be
D_M {\cal H}^M=0\, \label{DH} .  \ee Essentially what we have done
here is beginning with weight $w-1$ field equations $G^M$
in~\eqn{maxeom}, which obey $I\cdot G=0=D\cdot G$, but do not arise
directly from varying an action, we have formed equations of motion
${\cal H}^M$ which do follow from an action and obey 
\be X\cdot {\cal
  H}=0=D\cdot {\cal H}\, \label{DHX}.  
\ee 
Now the displayed Weyl invariant equations determine $\cal H$ from its 
middle slot.  To write explicitly a tractor formula for ${\cal H}^{M}$, we note that
the tractor 
\be \stackrel{\circ}{\cal
  G}{\!}^M= \Big\{G^M-\frac 1{(d+2w-2)^2} D^M X\cdot
G\Big\}\, , \ee 
has middle slot equaling the Proca equation
$\stackrel{\circ}{\cal G}{\hspace{-1.2mm}}^m={\cal G}^m$.
Then  we can produce a tractor obeying~\eqn{DHX}
of any desired weight $k+w-1$ from the quantity $D_{N}\Big[\sigma^{k}X^{[N}  (\stackrel{\circ}{\cal
  G}{\!}^{M]}-Y^{M]} X\cdot \stackrel{\circ}{\cal
  G})\Big]$ where the scale dependent, weight~$-1$ tractor~$Y^{M}$ obeys~$X\cdot Y=1$ and is constructed explicitly in 
 Appendix~\ref{projectors} . In particular (excepting two exceptional weights) the middle 
slot of this quantity is a non-zero multiple of a  power of $\sigma$ times ${\cal G}^{m}$.

The projection methods outlined in Appendix~\ref{projectors} can be used to express the
integrand $\sqrt{-g} \, {\cal L}$ of the action~\eqn{s2} as a tractor scalar \bea
\sigma^{d +2w-2}{\cal L} \; =\;& -&\frac 1{4(d+2w-2)^2}\,
\widehat {\cal F}_{MN} \widehat {\cal F}^{MN} \nn\\[2mm]
&-&\frac1{2\sigma^2}\, (w+1)(d+w-2)\, (I\cdot I)\; \wt V^M \wt V_M\, .
\eea 
Here the hat on the tractor Maxwell curvature denotes a tractor
covariant (but scale dependent) projection onto its middle slot as
explained in Appendix~\ref{projectors}. The first term is reminscent of the Maxwell
action, while the second reminds one of the Proca mass term.

\begin{figure}
\begin{center}
\epsfig{file=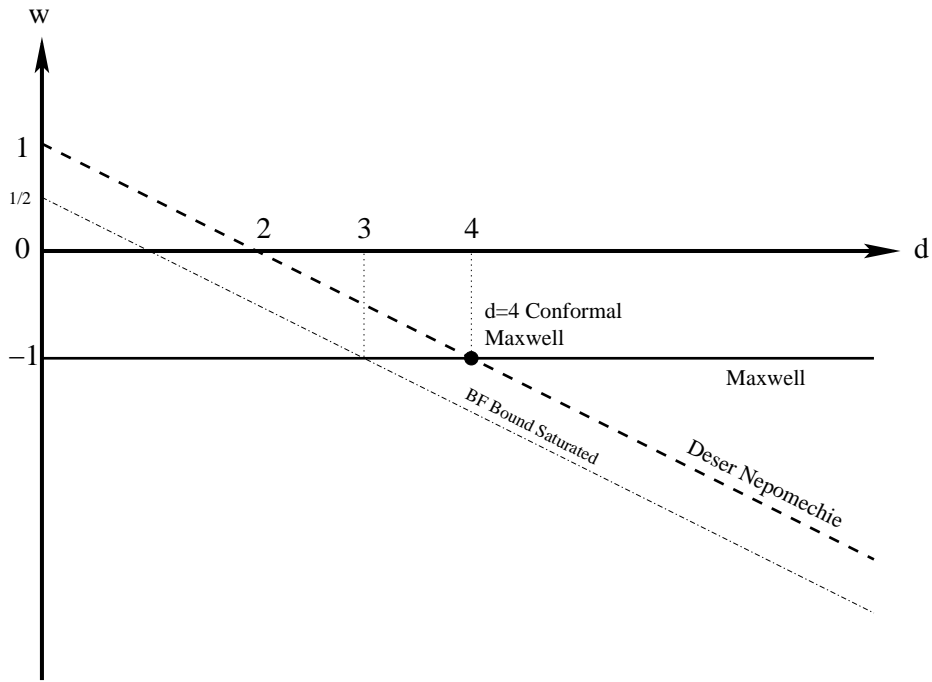,height=9cm}
\end{center}
\caption{A plot of the theories described by the tractor Maxwell system as a function of Weyl weight and dimension. Theories
saturating a vector Breitenlohner--Freedman bound appear at $w=\frac12-\frac d2$. \label{vectormass}}
\end{figure}

The various theories we have described using the single
equation~\eqn{maxeom} are plotted in Figure~\ref{vectormass}.

\subsection{On-Shell Approach}

Although, we have already given a complete and unified description of the Maxwell and Proca systems using tractors, 
it is useful to have an on-shell approach thanks to its simplicity and easy applicability to higher spin systems.

As prelude, we review the on-shell approach to the Proca equation in components. 
As explained above, the Proca system is described by the pair of equations
\be
\left\{
\begin{array}{l}
\Big(\Delta +{\textstyle\frac{ {\ts 2}\Rho}{{\ts  d}}}\, [w(w+d-1) -1]\Big) V_\mu=0\, ,\\[5mm]
\quad \nabla^\mu V_\mu = 0\, .
\end{array}
\right.\label{system}
\ee
The latter, divergence constraint, implies that the system describes $d-1$ propagating modes
which obey the Klein--Gordon equation. However, this is only true at generic values of the mass
(alias Weyl weight~$w$). For special values of $w$ there may be ``residual'' gauge invariance implying a 
reduction in degrees of freedom. Indeed, 
the (Maxwell) gauge transformation
\be
\delta V_\mu = \nabla_\mu \xi
\ee
leaves the divergence constraint invariant whenever the gauge parameter $\xi$ obeys
\be
\Delta \xi = 0.\label{5}
\ee
Then the identity
\be
[\Delta,\nabla_\mu]\xi = \frac{2\Rho}{d}\, (d-1) 
\nabla_\mu\xi
\ee
implies that the variation of the left-hand side of the Klein--Gordon equation equals
$
-m^2\nabla_\mu \xi\, ,
$
which vanishes whenever the Maxwell mass (defined in~\eqn{maxmass}) does.
{\it I.e.}, when $m^2=0$, there is a residual gauge invariance which removes an
additional degree of freedom so that the system of equations~\eqn{system}
describes $d-2$ propagating photon modes. (The divergence constraint is then
reinterpreted as the Lorentz choice of gauge.)

The above discussion was completely standard, but let us see if it can
be reproduced using tractors: The Proca system is now described by a
weight~$w$ tractor subject to \be D\cdot V=0\, .\label{1} \ee We can
fix the St\"uckelberg gauge invariance~\eqn{maxgauge} by setting \be
X\cdot V=0\, .\label{2} \ee Then the pair of equations~\eqn{system}
correspond to tractor equations \bea I\cdot D\ V^M&=&0\,
,\label{3}\\[2mm] I\cdot V \;\;&=&0\, .\label{4} \eea Our task now is
to search for residual gauge symmetries of the system of
equations~(\ref{1}-\ref{4}).  Clearly we should study the
transformation \be \delta V^M = D^M \xi\, , \ee where $\xi$ has weight
$w+1$.  Equation~\eqn{1} is trivially invariant under this gauge
transformation, but~\eqn{4} requires \be I\cdot D\ \xi = 0\, .  \ee
(This is the tractor analog of~\eqn{5}.) As a consequence,
equation~\eqn{3} is now invariant so the final condition is given by
varying~\eqn{2} which gives \be X\cdot D\ \xi = (d+2w)(w+1) \xi = 0\,
.  \ee At $w=-d/2$ special analysis is required as in the previous
section. For example here it follows that $D^M \xi \equiv 0$ which is
uninteresting.  At $w=-1$ we find a genuine residual gauge
invariance. This value of course implies $m^2=0$ in agreement with the
above component computation.

\section{Spin Two}

The massive spin two system is more subtle than its spin one Proca
relative. Its massless case corresponds to (linearized) gravitons so
introducing a general background would lead us to the non-linear
Einstein theory of gravitation. Also, it is generally accepted that
massive spin two systems cannot be coupled to general curved
backgrounds consistently
(see~\cite{Aragone:1979bm,Kobayashi:1978xd,Buchbinder:1999ar,Buchbinder:2000fy,Deser:2001dt}). However,
these difficulties can be circumvented in cosmological backgrounds
modulo subtleties--special tunings of the mass parameters lead to the
partially massless spin two
theory~\cite{Deser:1983tm,Higuchi:1986py,Deser:2001pe}.  In the
following we specialize to constant curvature theories\footnote{A
  study of non-minimal gravitational couplings for massive spin two
  fields yields consistent propagation in Einstein
  backgrounds~\cite{Buchbinder:1999ar,Buchbinder:2000fy}. Therefore we
  strongly suspect that there exist non-minimal tractor couplings that
  extend the results of this section to conformally Einstein
  metrics. We reserve this issue for future study.}. We first
introduce the various possible spin two theories---massive gravitons,
gravitons, partially massless spin two---using the simple on-shell
approach.

\subsection{On-Shell Approach}

Let us start directly with tractors;  the basic field is a weight $w$, symmetric rank two tractor $V^{MN}$.
As the spin two generalization of the spin one on-shell tractor equations~(\ref{1}-\ref{4}) we consider\footnote{The notation $D\cdot V^N$ is shorthand for $D_M
V^{MN}$.}
\bea
D\cdot V^N&=&0\, ,\label{DdotV}\\[1mm]
X \cdot V^N&=&0\, ,\label{XV} \\[1mm]
I\cdot D\ V^{MN}&=&0\, ,\label{IDV}\\[1mm]
I\cdot V^N &=& 0\, ,\label{IV}\\[1mm]
V_N^N&=&0\, .\label{trV}
\eea
The final trace relation is the only relation which is not a direct
analogue of a relation/equation in the spin one system of equations.
Written out in components these equations amount to the following triplet of
equations (as usual special attention is required at certain
weights, we will suppress discussion of this as the analysis is a
straightforward generalisation of that above for lower spins)
\bea
&&\Big(\Delta +\frac{2\Rho}{d}[w(w+d-1)-2]\Big)\, V_{\mu\nu}=0\, ,\\[2mm]
&&\; \; \nabla.V_{\nu}=0\, ,\\[2mm]
&&\quad V_\nu^\nu=0\, ,
\eea
where the symmetric tensor $V_{\mu\nu}$ sits in the middle slot of $V^{MN}$.
These are precisely the on-shell equations of motion for a massive spin two graviton.
Using the usual Pauli--Fierz definition of the spin two mass for which
\be
\Big(\Delta -\frac{4\Rho}{d}-m^2\Big)V_{\mu\nu}=0\, ,\label{PF2}
\ee
we find the spin two mass-Weyl weight relation
\be
m^2=\frac{2\Rho}{d}\Big[\Big(\frac{d-1}{2}\Big)^2-\Big(w+\frac{d-1}{2}\Big)^2   \Big]\, .\label{spin2mass}
\ee
This gives a Breitenlohner--Freedman bound
\be
m^2\geq \frac{2\Rho}{d}\Big(\frac{d-1}{2}\Big)^2\, .
\ee
Note that the expression~\eqn{spin2mass} factors as $m^2=-\frac{2\sRho}{d}\, w(d+w-1)$. We will find that the value $w=0$ at 
which the Pauli--Fierz mass term is absent, corresponds to the gauge invariant, massless graviton limit.

Let us now search for residual gauge invariances of the tractor equations of motion~(\ref{DdotV}-\ref{trV}).
First we consider transformations
\be
\delta V^{MN} = D^{(M}\xi^{N)}\, .\label{resid}
\ee 
where $\xi^N$ is a weight $w+1$ tractor vector. It is easy to see that equations~\eqn{DdotV},~\eqn{IDV},~\eqn{IV} and~\eqn{trV}
are invariant if
\be
I\cdot D \ \xi^N = 0 = I\cdot \xi = D\cdot \xi\, .\label{dots}
\ee
Invariance of~\eqn{XV}, however, requires
\be
(w+1)(d+2w) \xi^M + X^N D^M \xi_N=0\, .\label{theeq}
\ee 
Contracting this relation with $X_M$ yields
\be
2(w+1)(d+2w) X\cdot \xi=0\, .
\ee
It is not hard to verify that $D^{(M}\xi^{N)}\equiv 0$ at $w=-d/2$ so this value is uninteresting.
Hence either $w=-1$ or $X\cdot \xi=0$.  First we analyze the case $X\cdot \xi=0$. Using~\eqn{dots}
in conjunction with $X\cdot\xi=0$ we find
(using Appendix~\ref{tractor components}) that 
\be
X^N D^M \xi_N=-(d+2w)\xi^M\, .
\ee
Hence equation~\eqn{theeq} says
\be
w(d+2w) \xi^M=0\, .
\ee 
Thus we learn that~$\delta V^{MN}=D^{(M}\xi^{N)}$ is a residual gauge invariance\, 
of the on-shell equations~(\ref{DdotV}-\ref{trV}) at weight $w=0$ with~$I\cdot D\, \xi^M=0=I\cdot\xi=D\cdot \xi=X\cdot\xi$. 
This weight corresponds to vanishing Pauli--Fierz mass and in components the residual gauge invariance reads
\be
\delta V_{\mu\nu}=\nabla_{(\mu}\xi_{\nu)}\, ,\quad \mbox{ where }\; \nabla.\xi=0=\Big(\Delta-\frac {2\Rho}d\ (d-1)\Big)\xi_\mu\, .
\ee
This is a residual, linearized diffeomorphism so, as promised,  the $w=0$ theory describes constant curvature gravitons. 

Now let us turn to the other case $w=-1$. There the equation~\eqn{theeq} becomes
\be
X^N D^M\xi_N=0\, .
\ee
A solution to  this equation is given by setting
\be
\xi^M=D^M\alpha
\ee
where $\alpha$ has weight one. Comparing with~\eqn{resid} we see that $\alpha=\frac 1d X\cdot \xi$
and have therefore found a new residual gauge invariance
\be
\delta V^{MN} = \frac 1d D^M D^N X\cdot \xi\, . 
\ee
In components this transformation reads
\bea
\delta V_{\mu\nu} = (d-2)\Big\{\nabla_{\mu}\nabla_\nu + \frac{2\Rho}{d}\ g_{\mu\nu}\Big\} \xi^+ \, ,\quad\mbox{ where } (\Delta+2\Rho)\xi^+=0\, .
\eea
This double derivative, scalar gauge invariance is one that has been seen before -- it is the on-shell residual gauge invariance
of a partially massive spin two field~\cite{Deser:1983tm,Deser:2001pe}. The weight $w=-1$ corresponds to a mass
\be
m^2=\frac{2\Rho}{d}\, (d-2)\, .
\ee
In four dimensions, this gives the well-known result $m^2=2\Lambda/3$. This value gives a positive squared mass
in de Sitter space where $\Rho$ and $\Lambda$ are positive, and obeys (as does any real weight $w$) the Anti de Sitter
Breitenlohner--Freedman bound~\eqn{BFs}.

\subsection{Off-Shell Approach} 

Just as the Proca and Maxwell equations were unified in a single tractor equation
\be
I_M {\cal F}^{MN} =0\, ,
\ee
involving a choice of scale $\sigma$ and the tractor Maxwell curvature; massive gravitons,
gravitons and partially massless spin two theories can also be unified in a single equation
\be
I_R \Gamma^{RMN}=0\, .
\ee
Here $\Gamma^{RMN}$ are quantities that will be termed tractor Christoffel 
symbols. The reason for this name will soon be clear, but the reader is warned that we are  not asserting here that these are connection coefficients.
 Let us now explain this result in detail
by deriving it from first principles.

The model is described in terms of a weight $w$, rank two, symmetric tractor tensor.
The starting point, from which everything follows, is the gauge invariance
\be
\delta V^{MN}=D^{(M}\xi^{N)}\, .\label{2gauge}
\ee
There are various possibilities for the weight $w+1$ gauge parameter $\xi^M$. For example,
we could leave it unconstrained or ask it to satisfy relations built from $D^M$, $X^M$ and $I^M$.
The ``correct'' choice can be determined by comparison with the residual gauge symmetries discussed above and is\footnote{The model where
$D\cdot \xi=0$ is also interesting since it leads to the Weyl invariant spin two model for a trace-free symmetric rank two tensor
introduced by Deser and Nepomechie~\cite{Deser:1983tm}.}
\be
I\cdot \xi = 0\, .\label{paramc}
\ee
Hence in components
\be
\xi^M=\begin{pmatrix}\xi^+\\[1mm] \xi^m\\[1mm] \frac{\sf P}{d}\xi^+\end{pmatrix}
\ee
 and
\bea
\delta V^{++}&=&\;\;(d+2w)(w+1)\xi^+\, ,\nn\\[2mm]
\delta V^{m+}&=&\!\frac12 (d+2w) (w\xi^m+ \nabla^{m} \xi^+)\, ,\nn\\[1mm]
\delta V^{mn}&=&\;\; (d+2w)\Big(\nabla^{(m}\xi^{n)}+\frac{2\Rho}d \eta^{mn} \xi^+\Big)\, ,\label{2g}
\eea
while all other transformations are dependent on these ones. These are exactly the
gauge transformations of a St\"uckelberg approach to massive spin two excitations~\cite{Zinoviev:2001dt,Hallowell:2005np}.
Generically, they allow the auxiliary fields $V^{++}$ and $V^{m+}$ to be gauged away. 
At $w=0$, however, one obtains a massless graviton theory because the $\xi^+$ invariance
gauges away $V^{++}$ so $V^{m+}$ is inert and can be gauge invariantly set to zero, 
leaving  linearized diffeomorphisms \be\delta V_{\mu\nu}=d\nabla_{(\mu}\xi_{\nu)}\, .\ee 
At $w=-1$, the auxiliary $V^{++}$ is inert and can be set to zero, while $V^{m+}$ is also inert
so long as $\xi^+$ gauge transformations are accompanied by the compensating transformation
$\xi^m=\nabla^m\xi^+$ which yields the partially massless gauge transformation
\be
\delta V_{\mu\nu}=(d+2w)\Big(\nabla_{\mu}\nabla_\nu \xi^{+}+\frac{2\Rho}d \, g_{\mu\nu} \xi^+\Big)\, .
\ee
These results are consistent with the ones found in the above on-shell approach.

Having established the correct gauge transformations, we can now develop dynamics for our theory.
The gauge transformations $\delta V^{MN}$ in~\eqn{2gauge} obey a constraint
\be
D_M\delta V^{MN} = \frac12 D^N \, \delta V^M_M\, .
\ee
Therefore we impose the same constraint on our fields $V^{MN}$
\be
D\cdot V^N -\frac 12 D^N V^M_M=0\, .\label{Vc}
\ee
This constraint implies that the independent field content of the model is
the physical spin two field $V_{\mu\nu}$ along with the auxiliaries $V^{m+}$ and $V^{++}$.

To build field equations we form the {\em tractor Christoffel symbols}
\be 2\Gamma^{RMN}= D^M V^{NR}+D^N V^{MR} - D^R V^{MN}\, ,\label{Gids}
\ee 
which, as mentioned above, are not claimed here to be related to
connection coeffcicients for any connection.
Unlike the tractor Maxwell curvature, these are not gauge invariant but transform as
\be
\delta \Gamma^{RMN} = \frac 12 D^M D^N \xi^R\, .\label{Gamgauge}
\ee
Moreover they obey a trace and Thomas $D$ divergence identity
\be
\Gamma^{RM}{}_M=0=D_M \Gamma^{RMN}\, .\label{Cid}
\ee

The Christoffel gauge transformation~\eqn{Gamgauge} combined with the parameter
constraint~\eqn{paramc} imply that the tractor Christoffels contracted with $I^M$ are gauge invariant.
Hence we propose the gauge invariant equations of motion
\be
G^{MN}=-2 I_R \Gamma^{RMN}=0\, .\label{2eom}
\ee
These equations  are, of course, not all independent. 
In fact we expect them to obey relations corresponding to the field constraint~\eqn{Vc} as well as 
a Bianchi identity coming from the gauge invariance~\eqn{2gauge}.
This is indeed the case; the Christoffel identities~\eqn{Cid} imply
\be
G^M_M=0=D_M G^{MN}\, .
\ee
These are alone not sufficiently many relations, since we would predict a pair of tractor vector relations.
However, a simple computation shows that $G^{MN}$ contracted with $I^M$ is parallel to the Thomas $D$ operator
\be
I_M G^{MN}=D^N\!{\cal X}\, ,
\ee
with ${\cal X}=I_M I_N V^{MN}$. 

It is easy to compare our proposed equations of motion with those found above via an on-shell approach.
Expanding out~\eqn{2eom} gives
\be
I\cdot D\ V^{MN} - 2 D^{(M} I\cdot V^{N)}=0\, .
\ee
It can be checked explicitly that choosing a gauge where $V^{++}=V^{m+}=0$ and using the field equations
implies that $X\cdot V^N=I\cdot V^N=D\cdot V^N=0$. In turn $I\cdot D\ V^{MN}=0$ 
which yields the on-shell equations~(\ref{DdotV}-\ref{trV}). 

Instead of choosing a gauge however, one can also compute by {\it tour de force} the component expressions for $G^{MN}$
and verify that they reproduce the known St\"uckelberg equations of motion for massive spin two in constant curvature backgrounds.
We devote the remainder of this section to this calculation.

Our first step is to solve the field constraint~\eqn{Vc} for $(V^{--}, V^{m-},V^{+-})$ in terms of the independent
field components $(V^{mn}, V^{m+}, V^{++})$. The results are displayed in Table~\ref{sol_to_constraints}. We employ the symmetric
tensor algebra notation explained in Appendix~\ref{symmetric}. In this vein we have denoted
\be
V=V_{\mu\nu} dx^\mu dx^\nu\, \mbox{ and } V^\pm= V^\pm_\mu dx^\mu\, ,
\ee
and similarly for the components of $G^{MN}$.
Next we must compute various components of the tractor Christoffel symbols~\eqn{Gids}; the results are given in Table~\ref{tractor_christoffels}.

\newpage

\begin{landscape}
\begin{table}[h]
\begin{tabular}{|c|}
\hline \\[2mm]
$V^{+-} = \frac{1}{d(d+2w)}\left( [\Delta-(d+w)\Rho]V^{++}-(d+2w+2)\,{\bf div}\, V^+ +\frac{w^2+w+d+wd/2}{2}\, {\bf tr}\, V \right)$  \\[4mm] \hline 

\\[2mm]
$ V^{-}   =   \frac{1}{(d+w)(d+2w)} \Big ( [\Delta -\frac{\sRho}{d}(d(d+w)+2(w+1))] V^{+} -\frac{d+2w}{2}\, {\bf div}\, V+\frac{1}{d}\, {\bf grad}\, [\Delta -(d+w-2)\Rho]V^{++}$ \\ 
$-\frac{d+2w+2}{d}\, {\bf grad}\ {\bf div}\, V^+ +\frac{d(d+2w)+w(d+2w+2)}{4d}\, {\bf grad}\ {\bf tr}\,  V \Big )$ 
\\[4mm] \hline

\\[2mm]
$V^{--} = \frac{1} { d(d+w)(d+w-1)(d+2w)} \Big (  \frac{d(d+2w)}{2}\, {\bf div}^2 \,  V - \frac{1}{2}\{[(d+w)^2+w]\Delta + w(d+w)(d+w-1)\Rho\}\, {\bf tr}\,  V$   \\
$+ 2\{(w+1)\Delta + [(d+w)(d+w-1) +2(w+1)]\Rho\}\, {\bf div}\, V^+ -[\Delta^2+2\Rho\Delta - (d+w)(d+w-1)\Rho^2]V^{++} \Big ) $ \\[4mm] \hline

\end{tabular}
\caption{Solutions to the constraint $D.V^N= \frac{1}{2}D^NV$}
\label{sol_to_constraints}
\end{table}

\newpage

\begin{table}
\begin{tabular}{|c|}
\hline  \\[2mm]
$2\Gamma^{+++} = w(d+2w-2)V^{++}$    
\\[2mm] \hline

\\[2mm]
$2\Gamma^{++n} = (d+2w-2)(\nabla^nV^{++}-2V^{n+}) $
\\[4mm] \hline 

\\[2mm]
$2\Gamma^{r++}=-(d+2w-2)(\nabla^n V^{++}-2(w+1)V^{n+})$ 
\\[4mm] \hline

\\[2mm]
$2\Gamma^{+mn}=-(w+2)(d+2w-2)V^{mn}+2(d+2w-2)\nabla^{(m}V^{n)+}+\eta^{mn}2(d+2w-2)(\frac{\sRho}{d}V^{++}+V^{+-} )$
 \\[4mm] \hline

\\[2mm]
$2\Gamma^{r+n}= w(d+2w-2)V^{rn}+(d+2w-2)(\nabla^nV^{+r} - \nabla^rV^{+n})$ 
\\[4mm] \hline

\\[2mm]
$2\Gamma^{rmn}=(d+2w-2)(2\nabla^mV^{rn} - \nabla^rV^{mn} + 2\Rho^{mn}V^{+r} + 2 \eta^{mn}V^{-r})$
\\[4mm] \hline

\end{tabular}
\caption{Tractor Christoffels used to build equations of motion.}
\label{tractor_christoffels}
\end{table}

\end{landscape}

Just as for the tractor Maxwell system, some equations of motion involve higher derivatives that can be eliminated.  The field equation $G^{++}$ involves no higher derivatives while the higher derivatives in $G^{+}$ and $G$  can be eliminated by studying  linear combinations of field equations as well as their traces and divergences:
\bea
{\cal G}_0^{++} & =& G^{++}\, , \nn \\[2mm]
{\cal G}_0^{+}\   &=& G^{+} - \frac{1}{d+w}\, {\bf grad}\,  G^{++} \, ,\nn \\[2mm]
{\cal G}_0 \;&=& G - \frac{2}{d+2w-2}\,{\bf grad}\, G^{+}\nn\\[3mm]
&+& \frac{2(d+2w-1)}{w(d+w-1)(d+2w-2)}\, {\bf g}\, \Big({\bf div}G^{+} - \frac{1}{d+2w}\Delta G^{++}\Big)\, .\nn\\
\eea
Even though all the higher derivatives have been eliminated in $({\cal G}_0,{\cal G}_0^+,{\cal G}_0^{++})$, these equivalent equations of motion do not directly follow from an action. Bianchi identities implied by the gauge invariances~\eqn{2g} will serve as the guiding principle in our pursuit of equations of motion $({\cal G},{\cal G}^+,{\cal G}^{++})$ that do come directly from an action principle \be S=\frac 12\int\sqrt{-g}\Big(V^{++}{\cal G}^{++}+V_m^+ {\cal G}^{m+}
+V_{mn}{\cal G}^{mn}\Big)\, .\ee  
The required Bianchi identities are
\bea
-{\bf div}\, {\cal G} + \frac{1}{2}w{\cal G}^+ & =& 0 \, ,\nn \\[2mm]
\frac{2\Rho}{d}\,{\bf tr}\,{\cal G} - \frac{1}{2}\,{\bf div}\,{\cal G}^+ + (w+1){\cal G}^{++} &=& 0\, .
\eea
These are solved by the following combinations of field equations
\bea
{\cal G}^{++} &=& \frac{4\Rho}{dw(d+2w-2)} \Big [ \frac{2(d+w-2)(d-1)\Rho}{(w+1)}{\cal G}_o^{++} - (d+w)\,{\bf div}\, {\cal G}_o^{+} \Big ]\, ,\nn \\ [4mm]
{\cal G}^+ \ &=&\frac{-8(d+w-1)(d+w)\Rho}{dw(d+2w-2)}\,  {\cal G}_o^+ \, ,\nn \\[2mm]
{\cal G} \;\;&=& {\cal G}_o - \frac{1}{4}\,{\bf g}\,{\bf tr} \,{\cal G}_o \nn\\[2mm]
&-&\frac{\Rho}{(d+2w-2)}\,  \Big [ 1- \frac2w\, \Big(1-\frac{w+1}{d+2w} -\,\frac{(d-1)^2}{d}\, \Big)\Big ]\, {\bf g}\, {\cal G}_o^{++} \, .
\nn \\[2mm]
\eea
Explicit expressions for ${\cal G}^{++}$, ${\cal G}^{+}$, and ${\cal G}$ are provided in the accompanying Table~\ref{companytable}. 
They are gauge invariant, derive from an action principle and obey the above identities. We finish this section by showing they correctly
describe the massive spin two system in constant curvature along with its massless and partially massless limits.

Using the gauge invariance~\eqn{2g} the fields $V^{+}$ and $V^{++}$ can be gauged away (so long as $w\neq0,-1$) and the spin two equations of motion can be written in a simpler way in terms of the minimal covariant field content $V_{\mu\nu}$:
\be
{\cal G}_{\rm PF} \equiv  {\cal G}\lvert_{V^{+}=0, V^{++}=0} \ = G_{\rm Einstein} + G_{\rm mass}=0,\label{123}
\ee
where the linearized cosmological Einstein tensor is given by
\bea
G_{\rm Einstein} &=&\Big[ \Delta - \frac{4\Rho}{d}\, \Big]V - {\bf grad}\,{\bf div}V + \frac{1}{2}\, \Big[{\bf g}\,{\bf div}^2+{\bf grad}^2\,{\bf tr}\Big]V\nn\\[3mm]
 &-&\frac{1}{2}\ {\bf g}\,
\Big[\Delta+\frac{2\Rho}{d}(d-3)\Big]\,{\bf tr}\,  V \, .
\eea
and the Pauli--Fierz mass term~\cite{Pauli} is
\bea
G_{\rm mass} &=&  -m^2\Big[1- \frac{1}{2}\, {\bf g}\,{\bf tr}\, \Big]\,  V\, .
\eea
In these formul\ae\ the mass  $m^2$ is the same as given in~\eqn{spin2mass}.

Taking a  divergence of the cosmological Pauli--Fierz field equation~\eqn{123}, we learn the constraint:
\be
-m^2({\bf div}- {\bf grad}\,{\bf tr}) V = 0 \, . \\[2mm]
\ee
There is, a however, a further constraint obtained from the combination of a double divergence and trace of the field equation
\be
{\bf div}^2\, {\cal G}_{\rm PF} +\frac{m^2}{d-2}\,{\bf tr}\,  {\cal G}_{\rm PF} =\frac{d-1}{d-2}\, m^2\, [m^2-\frac{2\Rho}{d}(d-2)]\, {\bf tr}\, V=0
\ee
Hence when
\be
m^2\neq 0,\;\frac{2\Rho}{d}(d-2)\, ,
\ee
we immediately find
\be
{\bf tr}V = 0 = {\bf div}V\, .
\ee
These constraints imply that the field $V_{\mu\nu}$ describes the $(d+1)(d-2)/2$ degrees of freedom of a massive spin two excitation.
The same results can also be obtained from $ {\cal G}^{++} \lvert_{V^{+}=0,V^{++}=0}$ in conjunction with  ${\cal G}^{+} \lvert_{V^{+}=0,V^{++}=0}$.  
Plugging the constraints into~\eqn{123}, we recover~\eqn{PF2}--the massive on-shell spin two equation
\be
(\Delta - \frac{4\Rho}{d}-m^2)V = 0\, .
\ee
Finally, the special masses $m^2=0$ and $m^2=(2\Rho/d)(d-2)$ correspond to weights $w=0$ and $w=-1$, respectively.
In these cases, the above constraints become the respective Bianchi identities
\be
{\bf div} \, {\cal G}_{\rm PF}= 0\, ,\qquad \Big[{\bf div}^2+\frac{2\Rho}d \, {\rm tr}\Big] {\cal G}_{\rm PF}=0\, , 
\ee
corresponding to gauge invariances
\be
\delta V={\bf grad}\, \xi\, ,\qquad \delta V=\Big[{\bf grad}^2+\frac{2\Rho}{d}\Big]\, \xi^+\, .
\ee
These correspond to linearized diffeomorphisms and the partially massless gauge transformation of~\cite{Deser:1983tm}.
A detailed discussion of these theories is given in~\cite{Deser:2001pe}. This concludes our demonstration that the simple tractor
equations describe the cosmological spin two system.

\newpage

\begin{landscape}

\begin{table}
\begin{tabular}{|l|}
\hline \\[2mm]
${\cal G}^{++}= \frac{8(d+2w-1)(d-1)\sRho^2}{d^2w(w+1)} \square V^{++} - \frac{16(d+w-2)(d-1)\sRho^3}{d^2(w+1)}V^{++}  
-\frac{16(d-1)(d+w-1)\sRho^2}{d^2w}{\bf div}V^+ - \frac{2\sRho}{d} \square {\bf tr} V +\frac{2\sRho}{d}{\bf div}^2 V + \frac{4(d-1)(d+w-2)\sRho^2}{d^2}{\bf tr} V$ \nn \\[4mm]  
 
\hline \\[2mm] 
${\cal G}^+=-\frac{8(d+w-1)\sRho}{dw} \Big ( -\frac{2(d-1)\sRho}{d} {\bf grad} V^{++} -\frac{w}{2}[{\bf div} - {\bf grad}\,{\bf tr}]V+[\square-{\bf grad}\,{\bf div}]V^+  
+ \frac{4\sRho(d-1)}{d}V^+ \Big )$ \nn \\[4mm] 

\hline \\[2mm]
${\cal G}= G_{linear Einstein} + G_{PF}-\frac{4(d+w-1)\sRho}{d}[{\bf grad}-{\bf g}{\bf div}]V^+ -\frac{2\sRho}{d}{\bf g}\square V^{++} + \frac{2\sRho}{d}{\bf grad}^2 V^{++}  
+\frac{4(d-1)(d+w-2)\sRho^2}{d^2}{\bf g}V^{++}$  \nn \\[4mm] \hline
\end{tabular}
\caption{Equations of motions for spin 2 system.}\label{companytable}
\end{table}

\end{landscape}

\section{Arbitrary Spins}

\label{arbitrary}

The methods explicated in detail for spins $s\leq2$ can be applied also to higher spin systems,
which can be massive, massless or partially massless with gauge invariances ranging from a single
derivative on a tensor parameter (depth one) to $s$ derivatives on a scalar parameter (depth~$s$)~\cite{Deser:2001pe}.
(A useful review on the extensive higher spin literature is~\cite{Bekaert:2005vh}.)
Again these models are all unified by a single tractor equation of motion. We begin 
with an on-shell approach.

\label{s>2}

\subsection{On-Shell Approach}

In components, the on-shell field equations for massive higher spin $s$ fields in constant curvature backgrounds\footnote{We ignore the possibility
of mixed symmetry higher spin fields, although these should be simple to handle using our approach.} are given by
\bea
&&\Big(\Delta +\frac{2\Rho}{d}[w(w+d-1)-s]\Big)V_{\mu_1\cdots\mu_s}=0\, ,\\[2mm]
&&\; \; \nabla.V_{\mu_2\ldots \mu_s}=0\, ,\\[2mm]
&&\quad V^\rho_{\rho\mu_3\ldots\mu_s}=0\, ,\label{osh}
\eea
It is not difficult to verify that these follow from the tractor equations of motion
$$
D\cdot V^{M_2\cdots M_s}=
X \cdot V^{M_2\cdots M_s}=
I\cdot D\ V^{M_1\cdots M_s}=
I\cdot V^{M_2\cdots M_s} =
V_R^{RM_3\cdots M_s}=0\, ,
$$\be \ee
where $V^{M_1\cdots M_s}$ is a totally symmetric, rank~s, weight $w$ tractor tensor
and the mass-Weyl weight relation is the one suggested above
\be
\mu^2=\frac{2\Rho}d \Big[ \Big(\frac{d-1}{2}\Big)^2-\Big(w+\frac{d-1}{2}\Big)^2+s\Big]\, ,
\ee
defining the mass by~$\Delta V_{\mu_1\cdots \mu_s}=\mu^2\, V_{\mu_1\cdots \mu_s}$.
The Breitenlohner--Freedman bound is given in~\eqn{BFs}. 

Once again, these equations enjoy residual gauge invariances, but now at weights $w=s-2,s-3,\ldots,0,-1$.
In tractors these read simply
\be
\delta V^{M_1\ldots M_s}= D^{(M_1}\cdots D^{M_t}\xi^{M_{t+1}\ldots M_s)}\, ,
\ee
where $\xi^{M_1\ldots M_{s-t}}$ has weight $s-1$ and the parameter $t$ is called the depth
of a partially massless gauge transformation. Since these are on-shell residual transformations they
are also subject to
$$
X\cdot \xi^{M_1\ldots M_{s-t-1}}=
I\cdot \xi^{M_1\ldots M_{s-t-1}}=
D\cdot \xi^{M_1\ldots M_{s-t-1}}=0
$$
\be
I\cdot D\ \xi^{M_1\ldots M_{s-t}}=0=
\xi_R^{RM_1\ldots M_{s-t-2}}\, .
\ee
At depth $t$ the partially massless field $V^{M_1\ldots M_s}$ must have weight $w=s-t-1$
corresponding to masses
\be
\mu^2=-\frac{2\Rho}d \Big[ (s-t-1)(s-t-1+d)+t+1\Big]\, .
\ee 
These results reproduce those found earlier in~\cite{Deser:2001pe} by rather different methods.

\subsection{Off-Shell Approach}

The St\"uckelberg field content required to describe a massive spin~$s$ field in~$d$-dimensions
is equivalent to that of massless spin~$s$ field in~$d+1$ dimensions (see~\cite{Hallowell:2005np} for a detailed explanation\footnote{A related approach, in which AdS$_{d}$ higher spin fields are arranged in $O(d-1,2)$ multiplets with the aid of a compensating field can be found in~\cite{Vasiliev:2001wa}.}).
In~$d+1$ dimensions a massless spin~$s$ field is described by a totally symmetric rank~$s$ tensor
subject to the condition that its double trace vanishes. A counting of independent  field components therefore yields
\be
\begin{pmatrix}d+s\\ s\end{pmatrix}-\begin{pmatrix}d+s-4\\ s-4\end{pmatrix}\, .\label{numero}
\ee
The tractor description involves a weight $w$, totally symmetric, rank~$s$ tractor tensor $V^{M_1\ldots M_s}$ but again
field constraints are necessary.
Indeed, the same number of independent field components as in~\eqn{numero} solve the tractor field constraints\footnote{This follows from the binomial coefficient
identity
$$
\begin{pmatrix}d+s\\ s\end{pmatrix}-\begin{pmatrix}d+s-4\\ s-4\end{pmatrix}
=
\begin{pmatrix}d+s+1\\ s\end{pmatrix}-\begin{pmatrix}d+s\\ s-1\end{pmatrix}
+\begin{pmatrix}d+s-4\\ s-5\end{pmatrix}-\begin{pmatrix}d+s-3\\ s-4\end{pmatrix}
.
$$}
\be
D\cdot V^{M_2\ldots M_s}-\frac{s-1}{2}D^{(M_2} V_R^{M_3\ldots M_s)R}=0=V_{RS}^{RSM_5\ldots M_s}\, .\label{fc}
\ee
The first of these is consistent with our proposed gauge invariance
\be
\delta V^{M_1\ldots M_s}=D^{(M_1}\xi^{M_2\ldots M_s)}\, ,\label{sgauge}
\ee
where the parameter $\xi^{M_2\ldots M_s}$ is weight $w+1$ and obeys the parameter constraints
\be
I\cdot \xi^{M_1\ldots M_{s-2}} = 0 = \xi_R^{R M_3\ldots M_{s-1}}\, .
\ee
It is not difficult to write these out in components for the example of spin~3 (say) and check that they concur
with the general St\"uckelberg gauge transformations given in~\cite{Hallowell:2005np}. 

The field constraints~\eqn{fc} and gauge transformations~\eqn{sgauge} constitute the kinematics of our model. The dynamics
are determined by finding the gauge invariant higher spin generalization of the spin two equation of motion~\eqn{2eom}.
We conjecture this to be
\be
G^{M_1\ldots M_s}= I\cdot D\,  V^{M_1\ldots M_s} - s D^{(M_1}I\cdot V^{M_2\ldots M_s)}\, .
\ee
Its gauge invariance is trivially checked and it can also be rewritten in terms of higher spin tractor
Christoffel symbols\footnote{The generalized  Christoffel symbol approach to higher spins was pioneered in~\cite{de Wit:1979pe, Damour:1987vm}.}.  A proof requires verifying that this system of equations reproduces the on-shell ones~\eqn{osh};
we postpone it to  a forthcoming publication~\cite{GSW}. Gauge invariance, the matching of counting of field components
and our explicit $s\leq2$ computations are already strong evidence in favor of this conjecture.

\section{Gravity and Weyl Compensators}

\label{Compensate}

Another way to view the approach we advocate is in terms of a dilaton field viewed as
a  Weyl compensator. From a physical standpoint, much of what follows is standard material (see~\cite{Zumino,Deser0,Siegel}),
but it is still very useful to connect tractor and physical approaches. Consider the cosmological Einstein--Hilbert action
\be
S_{\rm EH}(g_{\mu\nu})=-\frac{1}{2\kappa^2}\int \sqrt{-g} \, \Big(R-2\lambda\kappa^{\frac{2}{1-d/2}}\Big)\, .
\ee
where $\Lambda=\lambda\kappa^{-\frac{2}{1-d/2}}$ so that $\lambda$ is dimensionless.
Setting 
\be
\sigma=\kappa^{\frac2{d-2}}\, ,
\ee
with $\kappa^{2}=8\pi G$ equaling the Newton constant,
we may rewrite the cosmological Einstein--Hilbert action in tractors as 
\be
S_{\rm EH}(g_{\mu\nu},\sigma)=\frac{d(d-1)}2\int \frac{\sqrt{-g}}{\sigma^d}\,  \Big(I\cdot I+\frac{2\lambda}{d(d-1)}\Big)=S_{\rm EH}(\Omega^2 g_{\mu\nu},\Omega\sigma)\,  .
\ee
We can add this Weyl invariant action to any of the matter actions we have discussed in previous sections and obtain a
Weyl invariant theory coupled dynamically to gravity\footnote{Modulo the usual issues involving lost gauge invariances and possible ghost excitations for higher spins in generally curved backgrounds.}.

The scale $\sigma^{1-d/2}$ plays the {\it r\^ole} of the dilaton $\varphi$,
\be
\varphi=\sigma^{1-\frac d2}\, .
\ee 
At arbitrary scales 
\be
S_{EH}(g_{\mu\nu},\varphi)=-\frac{4(d-1)}{(d-2)}\int \sqrt{-g}\, \Big(\,\frac{1}{2}[\nabla_\mu\varphi]^2+\frac{1}{8}\frac{d-2}{d-1}\, \Big[R\, -2\lambda\varphi^{\frac{2}{d-1}}\Big]\varphi^2\Big)\, ,
\ee
which is the Weyl invariant action for a conformally improved scalar with a conformally invariant potential. Fixing a gauge where $\sigma$ and therefore the Weyl compensator $\varphi$ is constant
\be
\varphi = \frac 1 \kappa\, ,
\ee
is often thought of as spontaneous breaking of Weyl invariance. The residual gauge  invariance of this choice of gauge ---constant rescalings of $\kappa$---
amounts to changes in the system of units. Then we obtain
\be
S_{\rm EH}(g_{\mu\nu},\kappa^{\frac{1}{d-2}})=-\frac1{\kappa^2}\int \!\sqrt{-g} \, (R-2\Lambda)\, ,
\ee
the standard cosmological Einstein--Hilbert action.

\section{Conclusions}

\label{conclude}

In this article we have given a general calculus for deriving physical theories based on Weyl invariance. The method is applicable
regardless whether the desired result is a massive, massless or partially massless theory. There are many possible applications of our ideas,
let us sketch just two of them here.

The theories exhibited in our work are not new, rather what we have developed is an efficient repackaging of massless, partially massless and 
massive theories as single Weyl invariant tractor theory. In itself, this is perhaps not so surprising, since a similar result can be obtained by radial dimensional reduction~\cite{SiegelB,Hallowell:2005np} of massless flat theories to obtain massless constant curvature ones in one dimension lower
(in fact those results essentially can be viewed as inhabiting the top and middle slots of the tractors used here). It would be very desirable to use our formalism to generate new theories or a deeper understanding of existing ones. The latter is clearly possible, given the large body literature on
conformal geometries which can be employed to study Weyl invariant theories. However, even the possibility of writing new theories is open;
let us give some ideas.

We have used Weyl invariance in this article to derive rigid conformal invariance of theories in fixed backgrounds. For example, In dimension~4
it has been shown that maximal depth partially massless theories are conformally invariant~\cite{Deser:2004ji}. Since these theories have masses corresponding to weights $w=-1$ equaling the value $w=1-d/2$ in $d=4$ dimensions where the Thomas $D$-operator has only a bottom slot,
we see that this earlier result is really  a consequence of Weyl invariance. This argument can be turned around, {\it i.e.} if we find conformally
invariant theories in a given background, we should look for new Weyl invariant theories underlying them. An example is odd dimensional vector theories.
In even dimensional de Sitter spaces, Maxwell's equations can be solved in terms of Bessel functions of half-integer index~\cite{Deser:2001pe}.
These can be expressed as a slowly varying wave envelope multiplying massless plane waves, which explains the lightlike propagation of these theories.
This is no longer true in odd dimensions, strangely enough. But at least in three dimensions there is a resolution. In flat three-dimensional space a topological mass term can be added to Maxwell vector theory yielding massive propagation~\cite{Deser:1982vy}. The same mechanism in de Sitter space
with appropriate tuning of the mass parameter leads to lightlike and even conformal propagation~\cite{Carlip:2008eq}. This  suggests an
underlying Weyl invariant theory. It ought be possible to find this theory using the tractor techniques developed here. 

Self-interacting, massive gravity theories, save possibly when viewed as effective theories, are plagued with ghost excitations~\cite{Boulware:1973my}.
However, we cannot resist remarking that the tractor formulation of massive spin two theories in terms of tractor Christoffel symbols
bears many similarities to its massless cousin. But massless spin two interactions are described by the Einstein--Hilbert action, moreover,
it is known how to systematically derive that theory from its linearized approximation~\cite{Deser1}. It is tempting to think that a non-linear, interacting
theory of a perturbed tractor metric $\eta^{MN}+V^{MN}$ might be calculable along these lines. This is rather speculative, but it might at least
lead to new insights into the problems of interacting higher spins.

\section*{Acknowledgments}
ARG is supported by Marsden Grant no.\ 06-UOA-029.  The article was
completed during ARG's participation in the programme ``Geometric
Partial Differential Equations'' at the Institute for Advanced Study,
Princeton.  A.W. is indebted to the University of Auckland for its
warm hospitality during several visits that progressed this article.

\appendix

\section{Compendium of Weyl Transformations}

In this appendix, we list some elementary Weyl transformations
that follow from the metric transformation 
\be
g_{\mu\nu} \mapsto \Omega^{2} g_{\mu\nu}\, .\label{Weyl}
\ee
Throughout, we denote
\be
\Upsilon_\mu=\Omega^{-1}\partial_\mu\Omega\, .
\ee
Firstly the volume form transforms as
\be
\sqrt{-g} \mapsto \Omega^d \sqrt{-g}\, .
\ee
Since the vielbein, Levi-Civita connection, and $rho$-tensor belong to the tractor connection, their transformations follow from (5) and (6). Explicitly,
 \bea
e_{\mu}{}^m \;&\mapsto& \Omega e_{\mu}{}^m , \nn\\[3mm]
\omega_\mu{}^m{}_n&\mapsto&\omega_\mu{}^m{}_n-\Upsilon^m e_{\mu n}+\Upsilon_n e_{\mu}{}^m\, ,\nn\\[3mm]
\Rho_{\mu n}\  &\mapsto&\Omega^{-1}\Big(\Rho_{\mu n}-\nabla_\mu \Upsilon_n +\Upsilon_n \Upsilon_\mu -\frac12 e_{\mu n} \Upsilon.\Upsilon\Big)\, .
\eea 
Similarly, the transformations of Cotton--York and Weyl tensors follow from the transformation of the tractor curvature (not to be confused with the
tractor Maxwell curvature of section~\ref{vectors})
\be
{\cal F}_{\mu\nu}= [{\cal D}_{\mu}, {\cal D}_{\nu}]  \mapsto  U{\cal F}_{\mu\nu} U^{-1} \,\label{curvfo}
\ee
with
\be
{\cal F}_{\mu\nu}=
\begin{pmatrix}
0&0&0\\[2mm]
C_{\mu\nu}{}^m&W_{\mu\nu}{}^m{}_n&0\\[2mm]
0&-C_{\mu\nu n}&0
\end{pmatrix}\, .
\ee
Here the Cotton--York tensor equals
\be
C_{\mu\nu}{}^m=\nabla_\mu \Rho_\nu{}^m-\nabla_\nu \Rho_\mu{}^m\, ,
\ee
and explicitly equation~\eqn{curvfo} says 
\bea
W_{\mu\nu}{}^m{}_n &\mapsto& W_{\mu\nu}{}^m{}_n\, , \nn \\[2mm]
C_{\mu\nu}{}^m\  &\mapsto& \Omega^{-1}\Big(C_{\mu\nu}{}^m -W_{\mu\nu}{}^m{}_n \Upsilon^n \Big)\, .
\eea
Transformations for scalars, vectors, and one forms of weight~$w$ are given by
\bea
f &\mapsto& \Omega^w f \, ,   \nn \\[2mm] 
v^{\mu} &\mapsto& \Omega^w v^{\mu}\, , \nn \\[2mm]
\omega_{\mu} &\mapsto& \Omega^w \omega_{\mu}  \, .
\eea
Covariant derivatives acting on these  transform as follows:
\bea
\nabla_\mu f\ &\mapsto& \Omega^w[(\nabla_\mu  + w \Upsilon_\mu) \,f ]\, ,\\[2mm]
\nabla_\mu v^\nu &\mapsto & \Omega^w[(\nabla_\mu  +w\Upsilon_\mu)\, v^\nu+ \Upsilon_\mu v^\nu - \Upsilon^\nu v_\mu +
\delta^\nu_\mu\ \Upsilon. v] \, ,\\[2mm]
\nabla_\mu \omega_\nu&\mapsto& \Omega^w[(\nabla_\mu  +w\Upsilon_\mu)\, \omega_\nu -\Upsilon_\mu \omega_\nu -\Upsilon_\mu \omega_\mu + g_{\mu\nu}\,  \Upsilon. \omega] \, .
\eea

\section{Tractor Component Expressions}

\label{tractor components}

In this Appendix we tabulate the component expressions for some of the more important tractor
quantities used in the text. Table~\ref{Tracod0} gives  the tractor covariant derivative acting on scalars, tractor vectors 
and rank two symmetric tractor tensors while Table~\ref{Tracod1} gives the tractor Laplacian on the same objects. Finally, Table~\ref{Tracod2} gives the Thomas
$D$-operator acting on scalars and tractor vectors.

\newpage
\begin{landscape}

\begin{table}[h]

\begin{tabular}{|c|}
\hline \\[3mm]
${\cal D} _{\mu}f = \nabla_{\mu} f$ 
 \\   [6mm] \hline

\\[3mm]
${\cal D}_\mu V^M=
\begin{pmatrix}
\nabla_\mu V^+ - V_\mu\\[2mm]
\nabla_\mu V^m +\Rho_\mu{}^m V^+ + e_\mu{}^m V^-\\[2mm]
\nabla_\mu V^- - \Rho_\mu{}^n V_n 
\end{pmatrix}$  
\\ [13mm] \hline

\\[3mm]
${\cal D}_\mu V^{(MN)}  =
\left(
\begin{array}{ccc}

\nabla_\mu V^{++}-2V_\mu{}^+ &
\nabla_\mu V^{+n}-V_\mu{}^n+ \Rho _\mu{}^nV^{++}+e_\mu{}^nV^{+-}  &
\nabla_\mu V^{+-}  -V_\mu{}^- - \Rho _{\mu r}V^{r+} \\[2mm]

{\rm Symm} &
\nabla_\mu V^{mn}+ 2 \Rho _\mu{}^{(m}V^{n)+}+2e_\mu{}^{(m}V^{n)-} &
\nabla_\mu V^{m-}+  \Rho _\mu{}^mV^{+-}+e_\mu{}^mV^{--}- \Rho _{\mu r}V^{rm} \\[2mm]
 
{\rm Symm} & 
{\rm Symm} &
\nabla_\mu V^{--}-2 \Rho _{\mu r}V^{r-}
\end{array}
\right) $
\\[15mm]  \hline

\end{tabular}
\caption{Tractor covariant derivative acting on scalars, tractor vectors, and rank two symmetric tractor tensors.}
\label{Tracod0}
\end{table}
\newpage

\begin{table}[h]
\begin{tabular}{|c|}
\hline \\[3mm]
${\cal D} ^2f = \Delta f $ 
\\ [6mm] \hline

\\[3mm]
${\cal D}^2V^M=
\left(
\begin{array}{c}
(\Delta- \Rho )V^+ -2\nabla^nV_n -dV^-\\[1mm]
\Delta V^m + 2 \Rho ^{mn}(\nabla_nV^+ - V_n)+\nabla^{n} \Rho V^+ +2 \nabla^mV^- \\[1mm]
(\Delta -  \Rho )V^- -2 \Rho ^{mn}\nabla_mV_n- \Rho ^{mn} \Rho _{mn}V^+ - V^n\nabla_n  \Rho 
\end{array}
\right) $
\\ [13mm]  

\hline \\[3mm]
${\cal D}^2 V^{(MN)} =$ 

$\left (  
{\tiny
\begin{tabular}{ c | c | c }

\multirow{3}{*} {$ \Delta V^{++} - 2(2\nabla. V^{+} - V_r{}^r + \Rho V^{++}+dV^{+-}) $}& 
$ \Delta V^{+n}-2\nabla.V^{n}- \Rho V^{+n}-4 \Rho_r^n V^{r+}$  &
$\Delta V^{+-}-2\nabla.V^{-}-2 \Rho V^{+-}-dV^{--} $\\

& $-(d+2)V^{n-}+ \Rho ^{sn} \nabla_sV^{++}+2\nabla^nV^{+-} $ & 
$+2 \Rho _{sr}V^{sr}- \Rho _r^s\nabla_sV^{+r}- \Rho _r^s \Rho _s^rV^{++}$ \\  

&$+\nabla_s( \Rho ^{sn}V^{++}) $  &
$-\nabla_s( \Rho _r{}^sV^{r+})$ \\ [3mm] \hline

\multirow{3}{*}{Symm}& 
$\Delta V^{mn}+2 \Rho ^{sm}\nabla_sV^{+n}+2 \Rho ^{sm} \Rho _s^nV^{++} $&
$ \Delta V^{m-}+ \Rho ^{sm}(\nabla_sV^{+-}-4V_s{}^--2 \Rho _{sr}V^{r+})$\\ 

 &+$4\nabla^nV^{m-}+4 \Rho ^{nm}V^{+-}+2\eta^{mn}V^{--} $   &
 $+2\nabla^mV^{--}- \Rho _r{}^s(\nabla_sV^{rm}+ \Rho _s^rV^{m+})- \Rho V^{m-}$\\
 
&$-4 \Rho _r{}^nV^{rm}+2\nabla_s( \Rho^{sm}V^{+n}) $& 
$+\nabla_s( \Rho ^{sm}V^{+-})-\nabla^s( \Rho _{sr}V^{rm})$ \\[3mm] \hline

 \multirow{2}{*}{Symm} &  &$\Delta V^{--}-2 \Rho _{rs}(\nabla^sV^{r-}+ \Rho ^{rs}V^{+-}- \Rho _t^sV^{tr})$  \\
 & Symm &$-2 \Rho V^{--}-2V^{r-}\nabla^s \Rho _{sr}$ \\
\end{tabular}}
\right)$ 
\\[20mm] \hline
\end{tabular} 

\caption{Tractor Laplacian acting on scalars, tractor vectors, and rank two index symmetric tractor tensors.}
\label{Tracod1}
\end{table}

\begin{table}[h]

\begin{tabular}{|c|}

\hline \\[3mm]
$
D^M f=
\begin{pmatrix}
(d+2w-2) w f\\[2mm]
(d+2w-2)\nabla^m f\\[2mm]
-(\Delta+w\Rho)f
\end{pmatrix} $ \\[15mm]

\hline \\[5mm]

$
D^MV^N \!=\!

 \left(\!\!
\scalebox{.95}{\mbox{$
 {\tiny
 \begin{array} {ccc}  
w(d+2w-2) V^+ &  w(d+2w-2)V^n  & w(d+2w-2)V^-   \\[3mm]
 (d+2w-2)(\nabla^mV^+ -V^m) &  (d+2w-2)(\nabla^mV^n+ \Rho ^{mn}V^++\delta^{mn}V^-) & (d+2w-2)(\nabla^mV^- - \Rho _{n}^m V^n)\\[3mm]
 -(\Delta+(w-1) \Rho )V^+ + 2\nabla.V + dV^- &
-(\Delta+w \Rho ) V^n - 2 \Rho ^n_m(\nabla^mV^+ - V^m)-V^+\nabla^n \Rho  -2 \nabla^nV^- & 
-(\Delta+(w-1)  \Rho )V^- +2 \Rho ^{mn}\nabla_mV_n+ \Rho ^{mn} \Rho _{mn} V^++ V^n\nabla_n  \Rho 
 \end{array}} $}}
\!\! \right) $ \\[15mm]
 
 \hline

\end{tabular}
\caption{Thomas D-operator acting on functions and tractor vectors.}\label{Tracod2}
\end{table}

\end{landscape}

\section{Projectors}

\label{projectors}

Equipped with a choice of scale $\sigma$, it is possible to convert any component expression 
into a tractor one using a projector technique. The expressions obtained this way are often
unwieldy, so that it is better to work directly in tractors from first principles. Nonetheless, we sketch here
a few details of the construction.

Given the  scale $\sigma$, we build the weight zero tractor vector 
\be
I^M=\frac1d\, D^M\sigma\, ,
\ee
from which we construct a null vector
\be
Y^M=\frac{1}{X\cdot I}\Big(I^M-\frac{I\cdot I}{2\, X\cdot I}\, X^M\Big)\, ,\qquad Y\cdot Y=0\, .
\ee
obeying
\be
Y\cdot X=1\, .
\ee

Armed with the null vectors $X^M$ and $Y^M$
we can now define the top, middle and bottom slots
of a tractor vector $V^M$ by 
\be V^+\equiv X\cdot V\ ,\qquad V^m\equiv
(V^M-Y^M X\cdot V-X^M Y\cdot V)\, , \qquad V^-\equiv Y\cdot V\, , 
\ee 
Tautologically then,
\be
Y^M=\begin{pmatrix}1\\ \ 0\ \\ 0\end{pmatrix}\, ,
\ee
and the tractor metric decomposes as a sum of projectors
\be
\eta^{MN}=X^M Y^N + \Pi^{MN} + Y^M X^N\, .
\ee
For an arbitrary tractor vector $V^M$ we denote
\be
\widehat V^M \equiv \Pi^M_N V^N = \begin{pmatrix}0\\ V^m\\ 0\end{pmatrix}\, ,
\ee
and similarly for any tractor tensor. This method allows us to extract the components 
of any tractor.

As a simple example, we can relate the usual Maxwell curvature to the tractor one ${\cal F}^{MN}=D^M V^N-D^N V^M$
(see also~\eqn{FF})
using projectors
\be
\widehat{\cal F}^{MN}=\Pi^M_R\Pi^N_S {\cal F}^{RS} = 
\begin{pmatrix}0&0&0\\[1mm]
0&(d+2w-2)F^{mn}&0\\[1mm] 0&0&0\end{pmatrix}\, .
\ee

\section{Symmetric Tensor Algebra}

\label{symmetric}

Computations involving symmetric tensors with high or many  different ranks are greatly facilitated
using the algebra of gradient, divergence, metric, trace and modified wave operators first introduced by Lichnerowicz~\cite{Lichnerowicz:1964zz} and
systemized in~\cite{Damour:1987vm,Hallowell:2005np,Hallowell:2007zb} (see~\cite{Labastida:1987kw,Vasiliev:1988xc,Duval,Duval1} for other studies). For completeness, we review the key formul\ae\ here.
The key idea is to write symmetric tensors in an index-free notation
using commuting coordinate differentials, so that a symmetric rank $s$ tensor $\varphi_{(\mu_1\ldots \mu_s)}$ becomes
\be
\Phi = \varphi_{\mu_1\ldots \mu_s} dx^{\mu_1}\cdots dx^{\mu_s}\, .
\ee 
In this algebra,  it is no longer forbidden to add tensors of different ranks.
Then there are seven distinguished operators mapping symmetric tensors to symmetric tensors:
\begin{enumerate}
\item[${\bf N}$\;] --Counts  the number of indices \be {\bf N}\ \Phi = s \Phi \, .\ee
\item[${\bf tr}$\ ] --Traces over  a pair of indices \be {\bf tr}\  \Phi = s(s-1)\varphi^\rho{}_{\rho\mu_3\ldots \mu_s} dx^{\mu_3}\cdots dx^{\mu_s}\, .\ee
\item[${\bf g}$\;] --Adds a pair of indices using the metric
\be {\bf g}\  \Phi = g_{\mu_1\mu_2}\varphi_{\rho\mu_3\ldots \mu_{s+2}} dx^{\mu_1}\cdots dx^{\mu_{s+2}}\, .\ee
\item[${\bf c}$\;] --The Casimir of the $sl(2)$ Lie algebra obeyed by the triplet $({\bf g,N}+\frac d2,{\bf tr})$ 
\be
{\bf c}={\bf g}\ {\bf tr} -{\bf N}({\bf N}+d-2)\, .
\ee
\item[${\bf div}$\ ] --The symmetrized divergence 
\be
{\bf div} \ \Phi = \nabla^\rho s\varphi_{\rho\mu_2\ldots \mu_s} dx^{\mu_2}\cdots dx^{\mu_s}\, .
\ee
\item[${\bf grad}$] --The symmetrized gradient
\be {\bf grad} \ \Phi = \nabla_{\mu_1}\varphi_{\mu_2\ldots \mu_{s+1}} dx^{\mu_1}\cdots dx^{\mu_{s+1}}\, .\ee
\item[$\square$\;] --The constant curvature Lichnerowicz wave operator
\be
\square=\Delta+\frac{2\Rho}{d}\, {\bf c}\, .
\ee
\end{enumerate}
The calculational advantage of these operators is the algebra they obey
\bea
[{\bf N},{\bf tr}]=-2{\bf tr}\, ,\quad
[{\bf N},{\bf div}]=-{\bf div}\, ,\quad
[{\bf N},{\bf grad}]={\bf grad}\, ,\quad
[{\bf N},{\bf g}]=2{\bf g}\, ,\nn
\eea
\bea
[{\bf tr},{\bf grad}]=2{\bf div}\, ,\quad
[{\bf tr},{\bf g}]=4{\bf N}+2d\, ,\quad [{\bf div,g}]=2{\bf grad}\, ,\nn\eea
\bea
[{\bf div,grad}]=\square-\frac{4\Rho}{d}\, {\bf c}\, .
 \eea
All other commutators vanish. In particular the Lichnerowicz wave operator is central!

\end{document}